\documentclass[universe,review,accept,moreauthors,pdftex]{Definitions/mdpi} 
\graphicspath{{Definitions/}}

\def\bc{{{BRITE}-Constellation}}
\def\br{{{BRITE}}}
\def\bt{{{BRITE-Toronto}}}
\def\hip{{{Hipparcos}}}
\def\ev{{Evris}}
\def\ed{{Eddington}}
\def\co{{CoRoT}}
\def\mo{{MOST}}
\def\te{{TESS}}
\def\wi{{WIRE}}
\def\sm{{SMEI}}
\def\cs{{Coriolis}}
\def\ga{{Gaia}}
\def\hs{{HST}}
\def\pr{{PRISMA}}
\def\st{{STARS}}
\def\ke{{Kepler}}

\def\pl{{Planck}}
\def\li{{Lisa}}
\def\po{{Plato}}
\def\cha{{Chandra}}
\def\c3{{CanX-3}}
\def\acir{{$\alpha$\,Cir}}
\def\kms{kms$^{-1}$}
\def\cd{cd$^{-1}$}

\newcommand{\bAur}{$\beta$\,Aur}
\newcommand{\epsAur}{$\varepsilon$\,Aur}
\newcommand{\zAur}{$\zeta$\,Aur}
\newcommand{\etaAur}{$\eta$\,Aur}
\newcommand{\tAur}{$\theta$\,Aur}
\newcommand{\nAur}{$\nu$\,Aur}
\newcommand{\iAur}{$\iota$\,Aur}

\newcommand{\bTau}{$\beta$\,Tau}
\newcommand{\kCet}{$\kappa^1$\,Cet}
\newcommand{\HR}{V711\,Tau}
\newcommand{\iqAur}{IQ\,Aur}

\usepackage{upgreek}
\firstpage{1} 
\pubvolume{7}
\issuenum{6}
\articlenumber{199}
\pubyear{2021}
\copyrightyear{2021}
\externaleditor{Academic Editors: Laszlo Szabados and Nikolay N. Samus} % For journal Automation, please change Academic Editor to "Communicated by"
\datereceived{17 May 2021} 
\dateaccepted{8 June 2021} 
\datepublished{16 June 2021} 
\hreflink{https://doi.org/10.3390/ \linebreak universe7060199} % If needed use \linebreak
\Title{Space Photometry with BRITE-Constellation \textsuperscript{\textsection}}

\TitleCitation{Space Photometry with BRITE-Constellation}

\definecolor{mypink1}{rgb}{0.858, 0.188, 0.478}

\definecolor{mygreen}{rgb}{0.7, 0.4, 0.8}

\definecolor{myblue}{rgb}{0.25, 0.41, 0.88}

\definecolor{mybgreen}{rgb}{0.5, 0.5, 1.0}

\definecolor{myyellow}{rgb}{1.0, 0.5, 1.0}

\Author{{Werner W. Weiss} $^{1,}$*\orcidA{},  Konstanze Zwintz $^{2}$\orcidB{}, Rainer Kuschnig $^{3}$\orcidC{}, Gerald Handler $^{4}$\orcidD{}, {Anthony F. J. Moffat} $^{5}$\orcidE{}, Dietrich Baade $^{6}$\orcidF{},
Dominic M. Bowman $^{7}$\orcidN{}, Thomas Granzer $^{8}$\orcidL{}, Thomas Kallinger $^{1}$\orcidP{}, Otto F. Koudelka $^{3}$, {Catherine C. Lovekin} $^{9}$\orcidI{}, Coralie Neiner $^{10}$\orcidR{}, Herbert Pablo $^{11}$\orcidT{}, Andrzej Pigulski $^{12}$\orcidM{}, Adam Popowicz $^{13}$\orcidG{}, Tahina Ramiaramanantsoa $^{14}$\orcidK{}, {Slavek M. Rucinski} $^{15}$\orcidQ{}, {Klaus G. Strassmeier} $^{8}$\orcidH{} and {Gregg A. Wade} $^{16}$\orcidO{} 
}
\AuthorNames{Werner W. Weiss, Konstanze Zwintz, Rainer Kuschnig, Gerald Handler, Anthony F. J. Moffat, Dietrich Baade, Dominic M. Bowman, Thomas Granzer, Thomas Kallinger, Otto F. Koudelka, Catherine C. Lovekin, Coralie Neiner, Herbert Pablo, Andrzej Pigulski, Adam Popowicz, Tahina Ramiaramanantsoa, Slavek M. Rucinski, Klaus G. Strassmeier and  Gregg A. Wade}

\AuthorCitation{Weiss, W.W.; Zwintz, K.; Kuschnig, R.; Handler, G.; Moffat, A.F.J.; Baade, D.; Bowman, D.M.; Granzer, T.; Kallinger, T.; Lovekin, C.C.; et~al.}

\address{%
$^{1}$ \quad Institute for Astrophysics, University of Vienna, A-1180 Vienna, Austria;  werner.weiss@univie.ac.at (W.W.W.); kallinger@astro.univie.ac.at (T.K.)\\ 
$^{2}$ \quad Institute for Astro- and Particle Physics, University of Innsbruck, A-6020 Innsbruck, Austria; konstanze.zwintz@uibk.ac.at\\
$^{3}$ \quad Institut für Kommunikationsnetze und Satellitenkommunikation, Technische Universität Graz, \mbox{A-8020 Graz, Austria;} rainer.kuschnig@tugraz.at (R.K.); koudelka@TUGraz.at (O.F.K.) \\
$^{4}$ \quad Nicolaus Copernicus Astronomical Center, Polish Academy of Sciences, 00-716 Warsaw, Poland; gerald@camk.edu.pl\\
$^{5}$ \quad Department of Physics, University of Montreal, Montreal, QC H3C 3J7, Canada; amoffat@sympatico.ca\\
$^{6}$ \quad European Southern Observatory (ESO), D-85748 Garching bei München, Germany; dbaade@eso.org\\ 
$^{7}$ \quad Institute of Astronomy, KU Leuven, B-3001 Leuven, Belgium; dominic.bowman@kuleuven.be\\
$^{8}$ \quad Department Cosmic Magnetic Fields, Leibniz Institut fuer Astrophysik Potsdam, D-14482 Potsdam, Germany; tgranzer@aip.de (T.G.); kstrassmeier@aip.de (K.G.S.)\\
$^{9}$ \quad Physics Department, Mount Allison University, Sackville, NB E4L 1E6, Canada; clovekin@mta.ca\\
$^{10}$\quad LESIA, Paris Observatory, PSL University, CNRS, Sorbonne Université, Université de Paris, \mbox{92195 Meudon, France}; coralie.neiner@obspm.fr\\
$^{11}$\quad American Association of Variable Star Observers (AAVSO), Cambridge, MA 02138, USA; hpablo@aavso.org\\
$^{12}$\quad Instytut Astronomiczny, Uniwersytet Wroc{\l}awski, 51-622 Wroc{\l}aw, Poland; pigulski@astro.uni.wroc.pl\\
$^{13}$\quad Department of Electronics, Electrical Engineering and Microelectronics, Silesian University of Technology, 44-100 Gliwice, Poland; Adam.Popowicz@polsl.pl\\
$^{14}$\quad School of Earth and Space Exploration, Arizona State University, Tempe, AZ 85287, USA; tahina@asu.edu\\
$^{15}$\quad Department of Astronomy and Astrophysics, University of Toronto, Toronto, ON M5S 3H4, Canada; rucinski@astro.utoronto.ca\\
$^{16}$\quad Department of Physics and Space Science, Royal Military College of Canada,  Kingston, ON K7k7b4, Canada; Gregg.Wade@rmc.ca
}
\corres{Correspondence: werner.weiss@univie.ac.at}
\thirdnote{\bc\ was designed, built, launched, and is operated and supported by the Austrian Research Promotion Agency (FFG), the University of Vienna, the Technical University of Graz, the University of Innsbruck, the Canadian Space Agency (CSA), the University of Toronto Institute for Aerospace Studies (UTIAS), the Foundation for Polish Science \& Technology (FNiTP MNiSW) and the National Science Centre~(NCN).}

\abstract{\bc\ is devoted to high-precision optical photometric monitoring of bright stars, distributed all over the Milky Way, in red and/or blue passbands. Photometry from space avoids the turbulent and absorbing terrestrial atmosphere and allows for very long and continuous observing runs with high time resolution and thus provides the data necessary for understanding various processes inside stars (e.g.,\ asteroseismology) and in their immediate environment. 
While the first astronomical observations from space focused on the spectral regions not accessible from the ground it soon became obvious around 1970 that avoiding the turbulent terrestrial atmosphere  significantly improved the accuracy of photometry and satellites explicitly dedicated to high-quality photometry were launched. A perfect example is \bc , which is the result of a very successful cooperation between Austria, Canada and Poland. Research highlights for targets distributed nearly over the entire HRD are presented, but focus primarily on massive and hot stars.}


\begin{document}

%============================================================================
\section{A Brief Flashback}    \label{flash}
%============================================================================
The first successful launch of a satellite in 1957 (Sputnik \cite{spu}) triggered a new era of astronomical observing techniques which  enormously expanded the research potential of astrophysics, mainly because the terrestrial atmosphere could be overcome. Consequently, the first space observations focused on spectral regions, which were not accessible from the ground. Eight years after { Sputnik}, Proton satellites observed cosmic  $\gamma$-rays (1965). { The Orbiting Astronomical Observatory (OAO \cite{oao}) from NASA were the first operational telescopes in space. After a power failure of OAO-1 right after the launch in 1966, OAO-2 was launched in 1968, and a follow-up, OAO-3 (Copernicus), was launched in 1972. These OAO-satellites provided a wealth of insight into the variability of stars and intricate details of the interstellar matter. The Astronomical Netherlands Satellite (ANS, 1974~\cite{ans}) conducted photometric observations of variable stars in the UV, followed by the NASA/ESA International Ultraviolet Explorer (IUE \cite{iue}) in 1978. } 

Already in the early days of space astronomy, the obvious scientific success accelerated the development of space instrumentation for all spectral ranges, taking advantage of avoiding a turbulent and absorbing atmosphere, not to mention clouds and a day/night rhythm. The desire to observe even fainter targets required the launch of space telescopes with increasing size---very similar to most ground based observatories. A still scientifically productive example is the amazing Hubble Space Telescope (\hs\ \cite{hst} ), launched in 1990. 

It needed nearly 30 years after the dawn of space telescopes for projects explicitly dedicated to ``simple'' stars to become a reality. A most prominent example is \hip\ \cite{hip} , an ESA mission with an aperture of 29 cm, launched in 1989, with the goal of determining high precision parallaxes of a large number of stars in our neighbourhood. The entire sky was scanned during three years, which resulted in up to 110 data points per star in the final  \hip\ and { Tycho} catalogues. The data have proven to be a treasure chest for detecting stellar variability (\citet{ka1} and many more). The follow-up ESA mission \ga\ (\citep{gai}), using a 145\,$\times$\,50\,cm telescope, was launched in 2013 and enormously increased  our knowledge about the 4D picture of our Galaxy.

Projects in the early stages of space telescopes focused on highly ranked targets devoted to the evolutionary aspects of galaxies, cosmology, interstellar nebulae and their role in stellar evolution, solar system planets and other hot topics. Monitoring classical variable or allegedly constant stars continuously over many hours, days or even months, was hindered due to the high pressure on telescope time. Fortunately, space telescopes need pointing and guiding equipment, usually provided by small auxiliary telescopes. One of the first star trackers, ``abused'' for stellar photometry, was the Fine Guidance Sensor (FGS) of the \hs , as described by \cite{ku9, ku2, zw1, we1, zw2} and in workshop proceedings of the Space Telescope Science Institute (\citet{ku1}).

Already in 1982, a first proposal for a space photometer dedicated to stellar variability and activity, \ev , was submitted to CNES (\citet{ma1}) and was developed further as a passenger instrument for the {{USSR-Mars94}} mission. It was intended to be active during the cruise time to Mars  (\citet{vu1}). However, launch of the {{Mars94}} mission was delayed to 1996 and ended in disaster, because of a rocket failure, which crashed {{Mars96}} in the Chilean Andes together with \ev . Fortunately, the experience gained during development of \ev\ was not lost: already in 1993, the French team had submitted  a larger follow-up seismology mission, \co\  (\citet{sc1, we2}), which was launched by ESA in 2006 and was active until 2014.

Asteroseismology experienced a boom towards the end of the last century, as it became obvious how much one can learn about stellar structure and evolution with this tool, as well as how one can test complex astrophysical concepts, with important implications for astrophysics in general. However, excellent data were necessary for such investigations, that is, data which cover as continuously as possible a long time span and with mmag accuracy or~better. 

\begin{figure}[H]

\includegraphics[width=12cm]{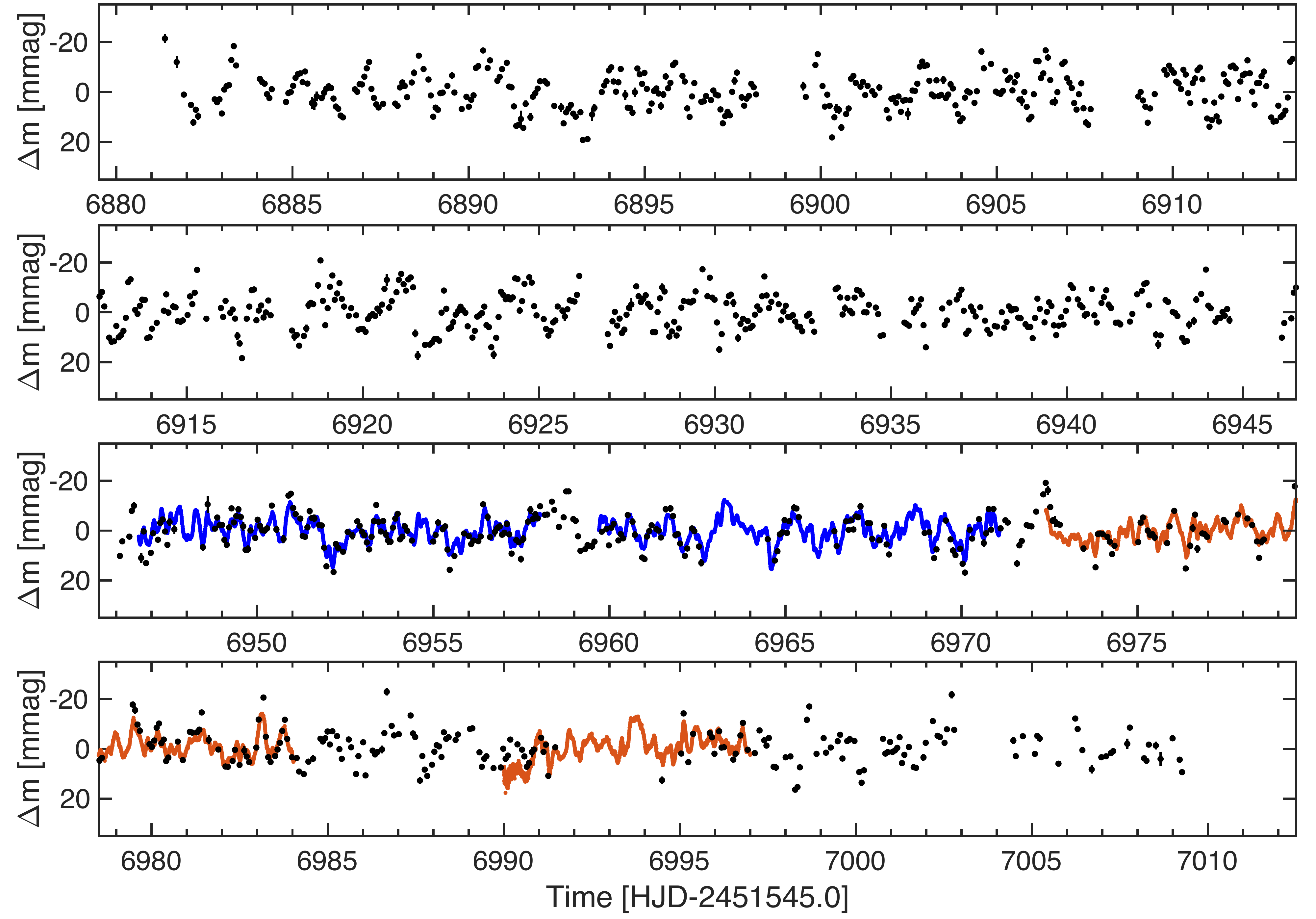}
     \caption{$\zeta$\,{Pup observed} by \br -Heweliusz, a component of \bc , (black dots, red filter)  and with \te\ in sector 7 (first line in blue) and in sector 8 (second line in red). %, using a CCD detector with no filter.
     }
     \label{figzpup}
\end{figure}

\begin{figure}[H]

\includegraphics[width=13cm]{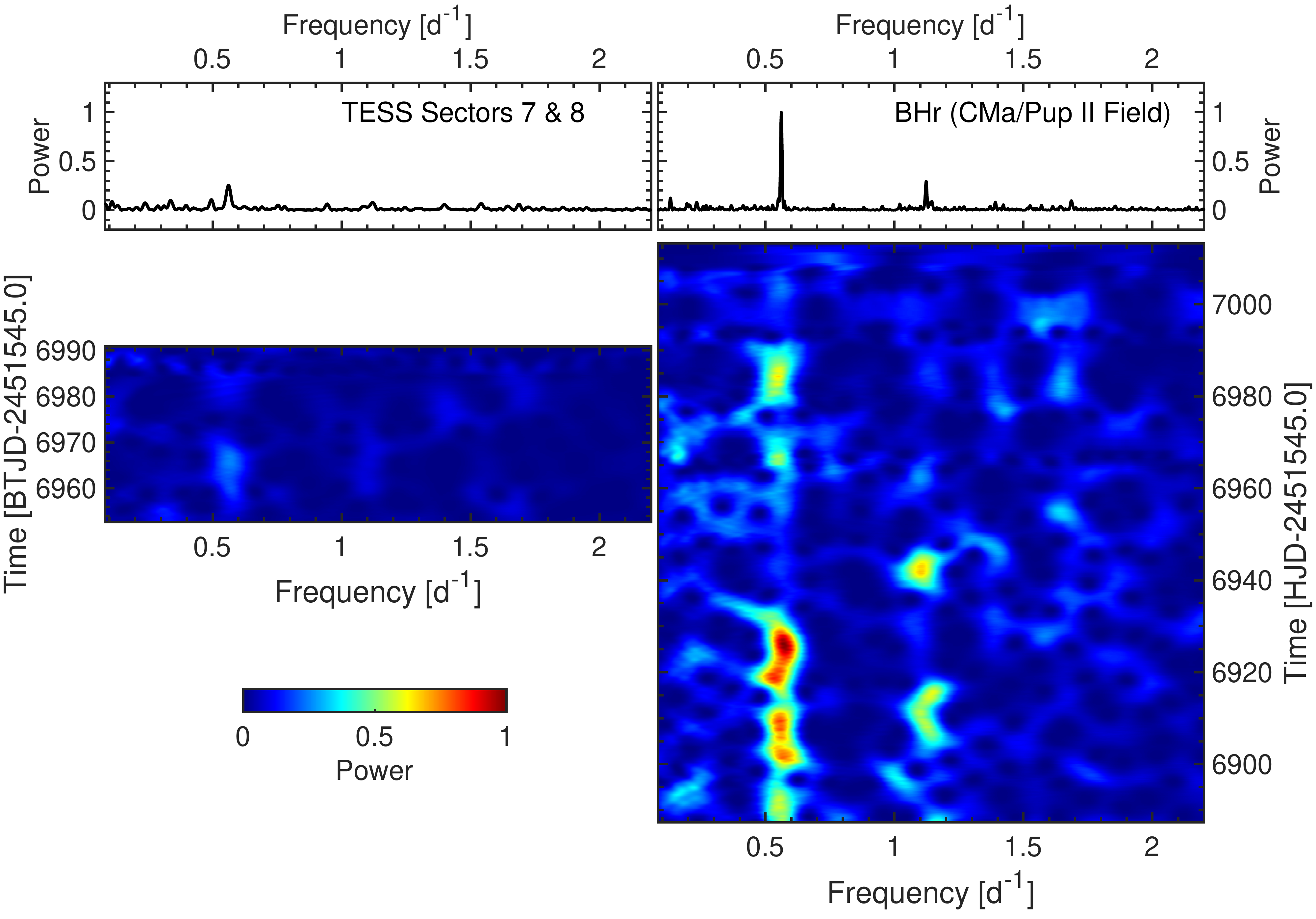}
     \caption{{Time dependent} frequency spectra of $\zeta$\,Pup obtained from data presented in Figure\,\ref{figzpup} (Ramiaramanantsoa et~al., in preparation) and with a sliding time window of 12\,days. The colour scale represents signal power normalized to the maximum power in the windowed discrete Fourier transforms of the BHr data.  }
     \label{figTEBR}
\end{figure}

 A textbook-like example is $\zeta$\,Pup (Figure\,\ref{figzpup}), which was observed simultaneously from space by two satellites and which illustrates the bonus of higher photometric accuracy (\te ) counter-weighted by longer data sets (\br ).
  Another example is \acir\ (\mbox{Section~\ref{s:alcir}} and {Figure 20}).
  The shorter \te\ run of $\zeta$\,Pup has broader Fourier peaks ($\approx$1.5 months vs. $\approx$4.5 months long data sets), but also shows many Fourier peaks, which appear to be less prominent in a longer run as is illustrated in the time resolved frequency analysis (lower part of Figure\,\ref{figTEBR}). Evidently, stochastic stellar variability dominates in the shorter run, while the 1.78\,d cyclic variability of the star is much more well-defined and well-covered in the longer observing run (see also Figure\,\ref{figzpup}).  The changing amplitude of the 1.78\,d signal indicates that the signal is likely due to (bright) spots of this O4If(n) type star that come and go. Sometimes the signal is not even there (see the BHr time-frequency diagram of \mbox{Figure\,\ref{figTEBR}}). 
 Whenever there are overlapping \te\ and BHr observations, they follow each other relatively well and have roughly the same amplitudes, except at the beginning of each subset of the TESS observations, which is due to systematics in TESS data around gaps.
  More about $\zeta$\,Pup is presented later in Section\,\ref{szPup}, including comments on the physics that is probably involved.
  %MDPI: Figures should be cited in numerical order 1234...  Current citation order is: 1,20,2,3,4,5,6,7,13,14,15,16,18,21,23,8,9,10,12,11,17,19,22, please revise --not applicable,  because description of the figures comes later in Sections 3 and 4.
  %MDPI: Figures should put below where it is first mentioned    --"mentioning" is not sufficient: explanation of the scientific background is the argument. 

%MDPI: in the image, Please change hyphen (-) to minus (−) in numbers --presently not possible to do   

In addition to  long continuous data sets, photometers operated in space have an advantage by avoiding day-time gaps in low-Earth orbits, inaccessibility of stars during certain seasons, and noise introduced by a turbulent atmosphere (\citet{we4}). 
The need for such data was subsequently boosted further by the discovery of exoplanets.  

\textls[-5]{The CNES mission \co\ provides a perfect example of a ``small'' satellite, which produced top science with a rather small budget, although nowhere nearly as small as that for \bc . Not surprisingly, the scientific community was interested in generating more such satellites, but competition with trendy space projects was intense, as is illustrated by the tortuous path \mbox{after \co .  \pr\  (\citet{le1})}}  was developed to extend \co\ and was accepted in 1993 as an ESA Horizon-2000 M2 project study, but finally lost the race in 2002 against the $\gamma$-ray satellite INTEGRAL.~The study team did not give up and produced a Horizon-2000 M3 proposal, \mbox{\st\  (\citet{jo1}}), but lost again in 2009 against the cosmic background explorer \pl .~The next attempt was \ed\ (\citet{fa1, ro1}), an ESA Flexi Mission, but the gravitational wave detector \li\  settled its problems for a planned launch in 2015, and consequently \ed\ had to step back. Later, the launch of LISA was delayed and is now scheduled for 2034. But finally,  PLATO was proposed in 2007 and selected in 2014 as an ESA Cosmic Vision mission (PLATO-Consortium {\citet{pla}}), driven by an exploding interest in exoplanets. Launch is scheduled for 2026. It took 20 years after \co\ till a follow-up, \po , finally was decided and about 30 years till-hopefully-first data will be available!

%MDPI: in the image, Please change hyphen (-) to minus (−) in numbers --presently not possible to do 

Outside Europe, similar efforts were also successful. Soon after the crash of \ev , an Announcement of Opportunity (AO) for Small Payloads was distributed in 1996 by the Canadian Space Agency (CSA), which was responded to in 1997 with a proposal for \mo\  (\citet{ru1}). 
This satellite was launched in 2003 as Canada’s first space telescope, and with an aperture of 15\,cm it was the smallest space telescope in orbit at that time. While designed only for a nominal lifetime of one year, it collected under the directorship of Jaymie Matthews (UBC) scientifically useful data till January 2018, that is, for more than 15 years.  Even after the CSA operations, funding ended in 2014, \mo\ was frequently activated for pay-per-view observers.

Paying tribute to the exploding interest in exoplanets after the detection of 51b\,Peg, NASA decided in 2001 to fund a space telescope, \ke , dedicated to the discovery of exoplanets (\citet{bo1}). At that time only 80 exoplanets were known, a number which increased dramatically after \ke 's launch in 2009. Reaction-wheel failures in 2012 and 2013 resulted in a modified mission, \ke -K2, which finally ended the mission in 2018, after the discovery of more than 2600 exoplanets and delivering an enormous amount of data for asteroseismology.

The Wide Field Infrared Explorer (\wi ) reminds one of the HST's Fine Guidance Sensors \textls[-15]{as auxiliary equipment with a potential for space photometry. \wi\ \mbox{(\citet{wir})}} was launched in 1999, but due to a premature ejection of the telescope cover, all cryogen quickly evaporated and made IR observation impossible. Fortunately, the star tracker was still working and contributed successfully to asteroseismology until the decommissioning of \wi\ in 2011. This substantially exceeded  the four months of the originally planned lifetime of the IR mission. Another mission producing photometric data for asteroseismology as a side-product to its main research goal is the Solar Mass Ejection Imager (\sm ) on board \cs\ (\citet{sme}), which was operational from 2003 to 2011 in a sun synchronous polar orbit with 102\,min period. 

The follow-up mission to \ke\  is \te , which was first discussed in 2005 and launched in 2018 by NASA, just after ending the \ke\ mission. \te\ (\mbox{\citet{ri1}} focuses on the stars brighter than those observed by \ke\ and the K2 follow-up, and it covers a sky area 400 times larger than that monitored by \ke . As an example of the relevance of \te\ data for asteroseismology we refer, for example, to \citet{cu1}, \mbox{\citet{an1}}, \citet{bow12} and \citet{bur7}.

More information about \hs , \ke , \ga\ and \te\ will be presented in dedicated chapters of this journal volume.

%============================================================================
\section{The Birth of {BRITE-Constellation}}  %%%%%%%%%%%%%%%%%%%%%%%%
%============================================================================

\begin{figure}[H]
%\center
     \includegraphics[width=10cm]{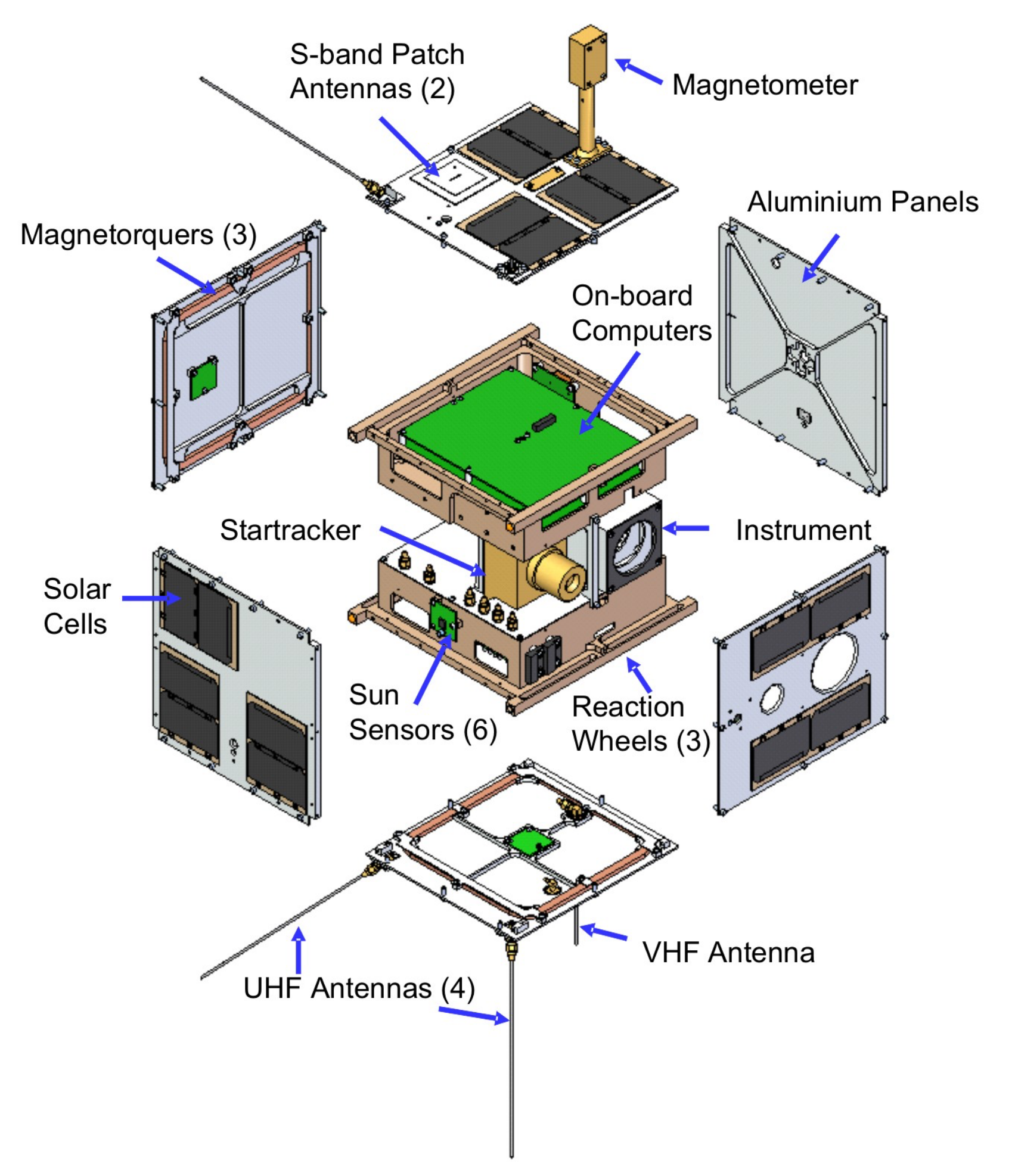}
     \caption{Basic structure of the \br\ satellites. Source: SFL.}
     \label{fig1}
\end{figure}

{ The development of \bc\ can be traced to the origins of the Canadian  microsatellite \mo\ \cite{ru1}, which was designed by Slavek Rucinski and Kieran Carroll (University of Toronto, UT), starting with construction in 1998, and successfully utilized by the team led by Jaymie Matthews (UBC) after launch in 2003 till 2014. Robert Zee (Manager of UT Space Flight Laboratory, SFL) wanted to continue the momentum created by the success of MOST and asked Rucinski in 2002 the non-trivial question, if nanosatellites could be of relevance for astronomy. One has to keep in mind that at that time nanosatellites were young and rarely utilized for research, with primary interest as an engineering experimentation exercise and looking down, not up, for Earth-atmosphere and -surface research. Nevertheless, a design concept for a single CANX-3 satellite was developed in 2004 by SFL and a small team of Canadian astronomers as a first fully three-axis stabilized satellite of 20 $\times$ 20 $\times$ 20\,cm size, containing a telescope with 3\,cm aperture (Figure\,\ref{fig1}).  }

Another root of origin is with Werner Weiss (University of Vienna, UoV) who was co-I of \ev , later of \co\ and also a member of the \mo\ team. The latter membership closed the loop to {\c3}. The failure of the \ev -launch contrasted dramatically with the anticipated research potential for asteroseismology, an expectation, which was later confirmed  by \co\ and \mo . Hence, the pressure to produce a space telescope optimised for bright stars grew.  Luckily, the Austrian Ministry of Science and Technology established in March 2005 a program for improving the infrastructure of Austrian Universities, to which UoV submitted a proposal for {\sc UniBRITE}. This was accepted in October 2005 and, one month later, {\sc UniBRITE} was ordered at SFL, based on their concept of {\c3}. 

A third root is with Otto Koudelka (Technical University Graz, TUG). The Austrian Space Agency (ASA) issued in 2005 a call for the 3rd Austrian Space Programme. Two nanosatellite proposals were in the queue: one of the Institute for Astrophysics (UoV), dedicated  to asteroseismology (\citet{we4}), and another from the Institute of Communication Networks and Satellite Communications (TUG), for developing and building a cubesat. ASA suggested to merge these initiatives, which resulted in a proposal with  Koudelka at TUG as the PI, and which was approved by ASA in 2006. This was the birth of the first satellite built in Graz (and Austria):  {\sc BRITE-Austria}, also called {\sc TUGSat-1}. The link to \mo\ is highlighted in a sentence of the proposal: {\sc BRITE-Austria} will extend and supplement the spectacularly successful Canadian microsatellite \mo\ into the domain of~nanosatellites.

As the Austrian \br 's were accepted for funding, Slavek Rucinski felt that Poland (his country of birth) with its rapidly improving economy should join. When the Canadian part of the project appeared to be in limbo due to CSA dragging its heels regarding funding (from 2006 until 2011), he started pushing his colleagues and former students in Poland to follow the Austrian example. Aleksander Schwarzenberg-Czerny (Copernicus Astronomical Center, Warsaw, CAMK and former PhD student of Rucinski) was able to obtain funding for two \br\ satellites at the end of 2009 and, hence, he provides the fourth root of \bc . 

The pressure on the Canadian Space Agency (CSA) increased considerably, after Austria funded {\sc UniBRITE} and {\sc BRITE-Austria}, and after Poland funded {\sc BRITE-Heweliusz} and {\sc BRITE-Lem}. Finally, CSA  accepted in 2011 the two Canadian \br 's (former \c3): {\sc BRITE-Toronto} and {\sc BRITE-Montreal}.

In this way, \bc\  was born with six satellites.

%============================================================================
\section{BRITE-Constellation}   
%============================================================================

The goal of \bc\ \cite{web1} was to provide high-precision photometric monitoring of very bright ($\lesssim$4 mag) stars in two optical wavelength bands (colours), that is, blue and red, and for up to six months, the maximum feasible time in an affordable low-Earth orbit. Various concepts have been discussed and finally a single telescope, optimized for a given passband was chosen with no moving elements on board, thus reducing risk, but which required one spacecraft per filter.

The proceedings of the First \br\ Workshop \cite{zw3} provide an overview to the technical and scientific issues, which were discussed and decided before launch in 2013. The situation of the six (five active) components of \bc\ after launch (\mbox{Table \ref{tab1}}) are described in \citet{we3, de1, ko1} and  various aspects of \br -data reduction in \mbox{\citet{pab1,po1,po2}}.

% start a new page without indent 4.6cm
%\clearpage
\end{paracol}
\nointerlineskip
\begin{specialtable}[H]
\widetable
\caption{Launch and orbital information for the \br\ nanosats. {\sc BRITE-Montr\'eal} did not separate from the launch vehicle and is not operational. The red filter covers 550--700\,nm, and the blue filter 400--450\,nm. }
%\center
\setlength{\cellWidtha}{\columnwidth/7-2\tabcolsep-0.2in}
\setlength{\cellWidthb}{\columnwidth/7-2\tabcolsep+1in}
\setlength{\cellWidthc}{\columnwidth/7-2\tabcolsep-0.3in}
\setlength{\cellWidthd}{\columnwidth/7-2\tabcolsep-0.4in}
\setlength{\cellWidthe}{\columnwidth/7-2\tabcolsep+0.0in}
\setlength{\cellWidthf}{\columnwidth/7-2\tabcolsep-0.2in}
\setlength{\cellWidthg}{\columnwidth/7-2\tabcolsep+0.10in}
\scalebox{1}[1]{\begin{tabularx}{\columnwidth}{
>{\PreserveBackslash\centering}m{\cellWidtha}
>{\PreserveBackslash\centering}m{\cellWidthb}
>{\PreserveBackslash\centering}m{\cellWidthc}
>{\PreserveBackslash\centering}m{\cellWidthd}
>{\PreserveBackslash\centering}m{\cellWidthe}
>{\PreserveBackslash\centering}m{\cellWidthf}
>{\PreserveBackslash\centering}m{\cellWidthg}}
\toprule
\textbf{Owner}                &  \textbf{Name}   & \textbf{Filter} & \textbf{ID}     & \textbf{Launch Date}  &          \textbf{Orbit (km)}          & \textbf{Period (min)}   \\
%                               &            &          &          &                      &            \textbf{}           &   \textbf{}     \\
\midrule
%\multicolumn{3}{c|}{Owner: Austria} & & &    \\
Austria & {\sc UniBRITE }      & red     & UBr  &25 Feb. 2013 & $781\times 766$ & 100.37 \\  
            & {\sc BRITE-Austria }      & blue   & BAb  &25 Feb. 2013 & $781\times 766$ & 100.36 \\
% \multicolumn{3}{c|}{Owner: Poland} & & &   \\
Poland & {\sc BRITE-Heweliusz}  & red     & BHr  &19 Aug. 2014 &    $612\times 640$   &   97.10 \\
            &{\sc BRITE-Lem }          & blue    & BLb  &21 Nov. 2013 & $600\times 900$ &  99.57 \\
%\multicolumn{3}{c|}{Owner: Canada} & & &   \\
Canada & {\sc BRITE-Toronto}      & red     &  BTr   &19 June 2014 & $629\times 577$&   98.24 \\
             &{\sc  BRITE-Montr\'eal}& {blue}&  &{19 June 2014} &                    &{n/a} \\          
\bottomrule
\end{tabularx}}
\label{tab1}
\end{specialtable}
\begin{paracol}{2}
%\linenumbers
\switchcolumn
%MDPI: Is the italic necessary? If not, please remove it.  --done

{ {Conferences were organised almost every year to discuss updates and new aspects of the mission. Most importantly, they allowed for vivid scientific discussions which helped to shape the focus of \bc . The first science conference took place in 2015 in Gda\'nsk, Poland, one year later in Innsbruck \cite{zw4}, and in 2017 at Lac\,Taureau, Canada.}}
The conference in Vienna ``Stars and their Variability, Observed from Space-Celebrating the 5th Anniversary of \bc '' in August 2019 provided the most recent status report \cite{proc1}.

\subsection{Instrumentation}

The \br\ instruments consist of a multi-lens telescope with an aperture of 3\,cm, optimised for the red (550--700\,nm) or the blue (400--450\,nm) wavelength range (\mbox{Figure\,\ref{fig2}}, red design). The unvignetted field of view (FOV) is about 24$^\circ$ in diameter and the optics were chosen to provide slightly out-of-focus stellar images for improved S/N, an experience acquired from MOST.  For the two Austrian, the two Canadian and the blue Polish instruments a 5-lens system was developed. The red Polish instrument (BHr) has a four-lens design, which results in a shorter telescope, but with a smaller FOV of 20$^\circ$. 
A baffle in front of each telescope reduces off-axis stray-light from bright sources, including the Sun, Moon and Earth. 
\begin{figure}[H]
%\center
     \includegraphics[width=11cm]{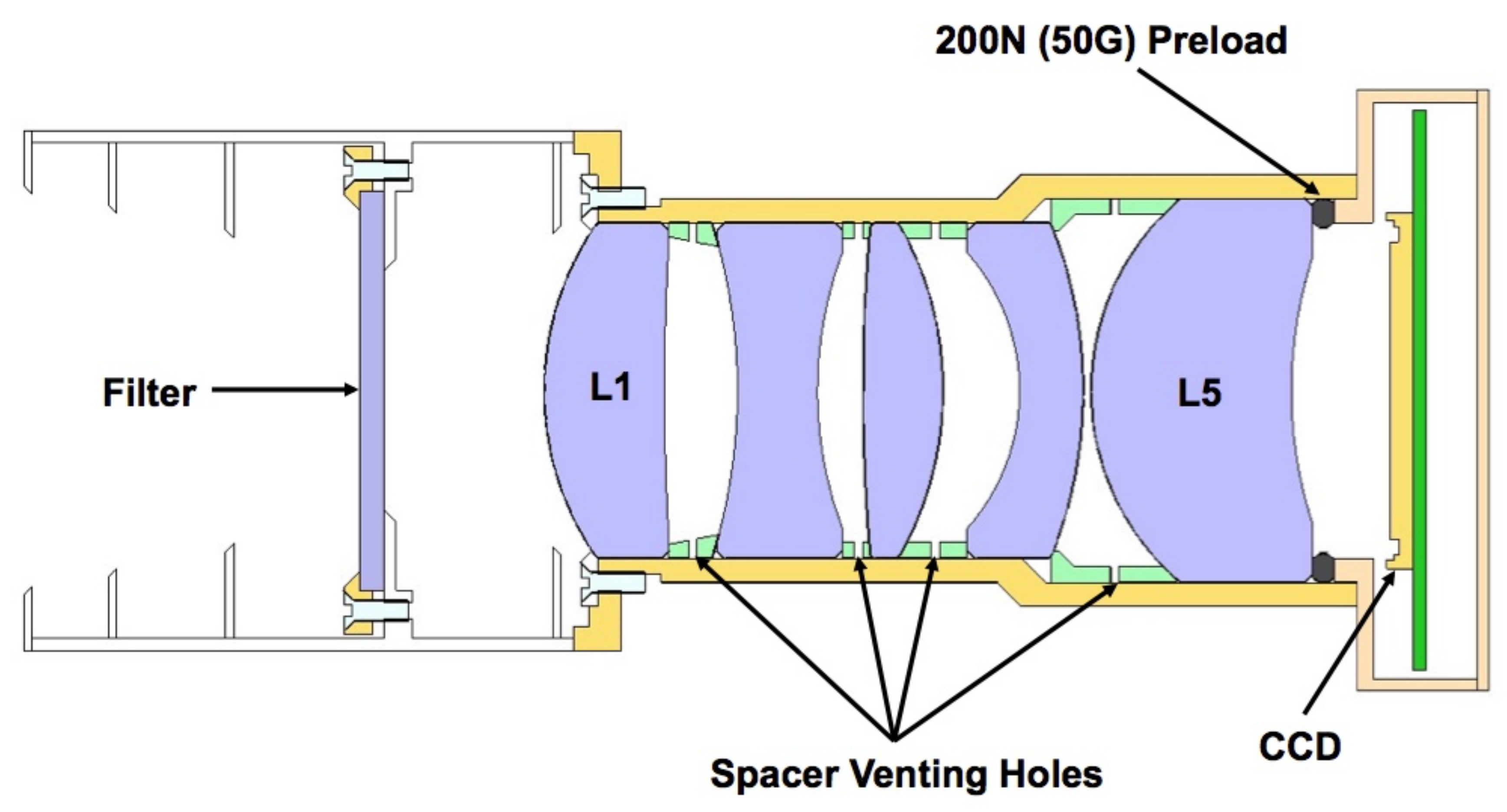}
     \caption{Camera scheme for the {\sc UniBRITE } and \bt\ red instruments with a nearly vignetting free field-of-view of 24$^\circ$. The red or blue filters are placed at the entrance pupil of the 5-lens camera to assure a constant filter function over the entire FOV. The blue cameras have slightly modified lens radii and separations to optimise the image quality for the needed wavelength range. Source: SFL.}
     \label{fig2}
\end{figure}

The same interline frame-transfer CCDs, a Kodak KAI 11002-M (4048 $\times$ 2672 pixels and 9 $\upmu$m pixel size) chip, are used for each \br . This is an off-the-shelf product which includes all read-out electronics and preamplifiers on a header board behind the chip.  Attractive features, besides the modest price, is the low dark current at high temperatures (0$^\circ$--30$^\circ$\,C), which allows one to avoid a cooling system, and a low read-out noise and power consumption. This CCD has been successfully used on the ground in SBIG Cameras, but never in the radiation environment of space.
In order to avoid pixel saturation, the CCD is positioned out-of-focus, which, together with the optical design, results in about 8-pixel-wide on-axis stellar images. Off-center images have a more complex shape, as is shown in Figure\,\ref{fig3}. The scale is about 27$'$$'$--30$'$$'$ per pixel, increasing slightly towards the edge due to image distortion.

\begin{figure}[H]

     \includegraphics[width=9cm]{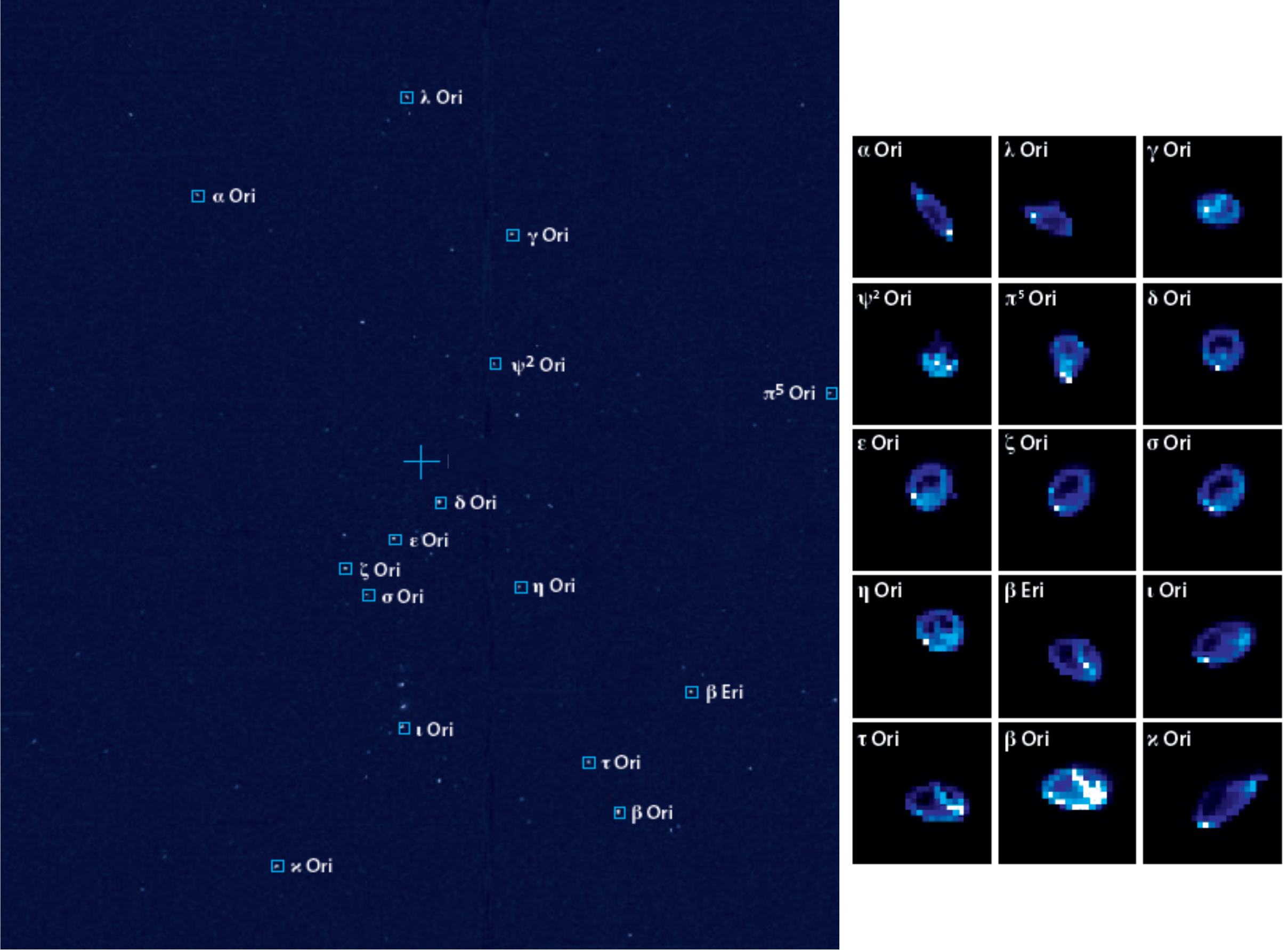}
     \caption{Full frame image of the Orion field taken with UniBRITE (UBr) in December 2013. The stars which have been selected for photometric time series are indicated. Subframes (24 $\times$ 24 pixels) which contain a full PSF, were stored in memory for a later download to the ground. Typical subframes in the center and close to the edge of the field are presented in the right panel.}
     \label{fig3}
\end{figure}

\subsection{Photometry and Data Processing}\label{sec32}

The \br\ mission requirements were set in 2005 such that the instruments shall observe a selected star-field for at least 15 min per orbit. Outside that time interval, scattered light from the Earth and Sun would be encountered. Data from up to 15 stars per field shall be collected for up to 100 days. 
In reality the \br\ satellites typically collect data from 24 to 28 stars during 20 to 40 min per orbit over a time base of about 160 days. The exposure times typically vary between 1 and 5 s and every 21 s subframes were read out. All functioning \br\ satellites were launched into low-earth polar orbits with periods close to 100 min.

Target fields can be occulted by the Earth during part of the orbit. After the field becomes visible again and its distance from the earth-limb exceeds a critical angle, the Attitude Control System (ACS) of the satellite re-points the star field in the camera FOV. To obtain top-quality photometry, the ACS must assure stable pointing of the PSF during the entire observing run close to the same pixels (flat-field exposures are not possible), which typically is achieved within 1.5$'$\,rms ($\approx$3 pixels). An example of such a photometric cadence is show in Figure\,\ref{fig4}. Whenever possible, a satellite setup was chosen, allowing one to observe a second field during an orbit, when the first field was invisible for the satellite. 

\begin{figure}[H]
%\center
     \includegraphics[width=13.5cm]{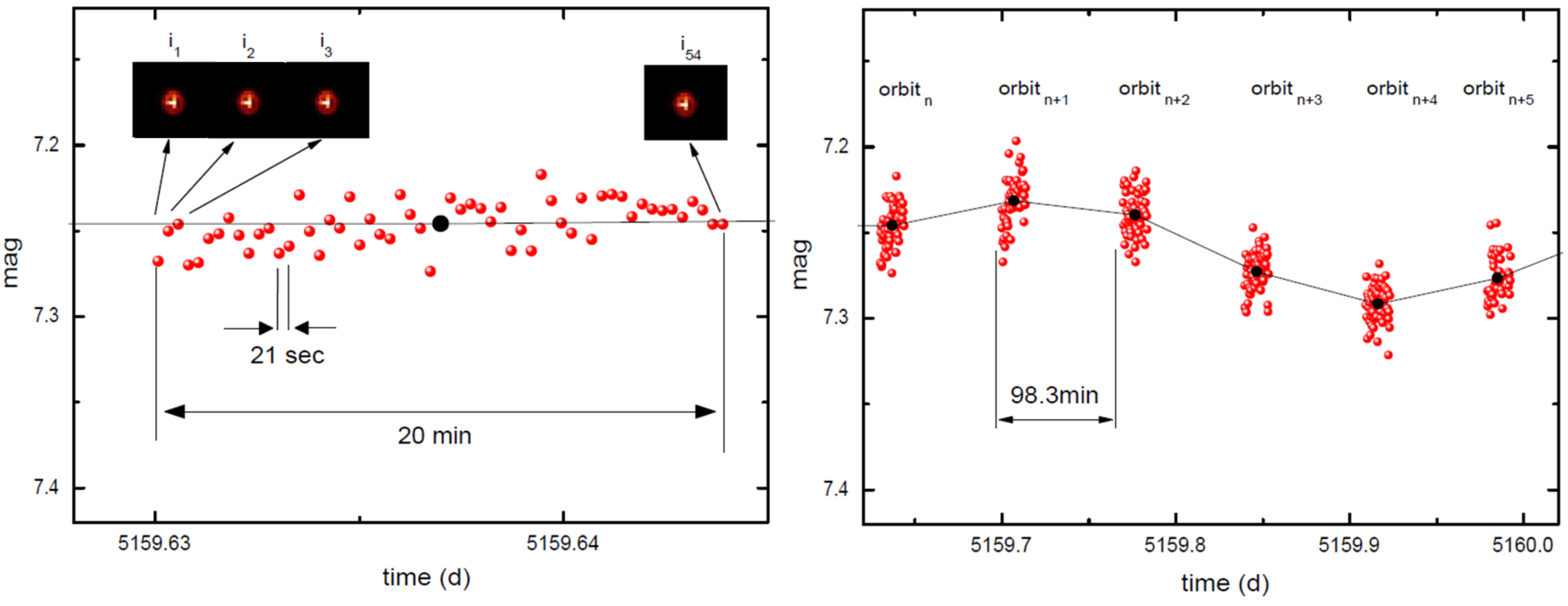}
     \caption{{A typical} photometric sequence of 44\,Cyg (\citet{zwintz1}), observed by \br -Toronto in the field during a single orbit (\textbf{left}) and the data sequence during six consecutive orbits (\textbf{right}), indicating intrinsic light variations. }
     \label{fig4}
\end{figure}
%MDPI: Please replace with a sharper image. --it's the best we can do

After the first \br\ satellites (UBr and BAb) were launched and the first images were recorded, features appeared which were not present in the laboratory: pixels and even entire columns with increased dark (thermal) signal, that is, ``hot pixels'' and ``warm columns'' (Figure\,\ref{fig5}). These flaws were distributed over a significant fraction of the CCD, in the FOV as well as outside with no light access. The defects became stronger during successive weeks, even at the same CCD operating temperature. The signal of ``warm columns'' ranged from 100 to 500\,ADUs above nominal background. One ADU corresponds at 20$^\circ$\,C to 3.2 detected electrons. For a ``hot'' pixel the  signal (even without illumination) is more than 100\,ADUs above median background and it can  even get close to saturation ($\approx$1200\,ADUs). 
All \br\ satellites suffer from these radiation defects, believed to arise mainly from proton collisions, which accumulate over time and adversely affect the data reduction and quality.\vspace{-6pt} 

\begin{figure}[H]
%\center
     \includegraphics[width=12cm]{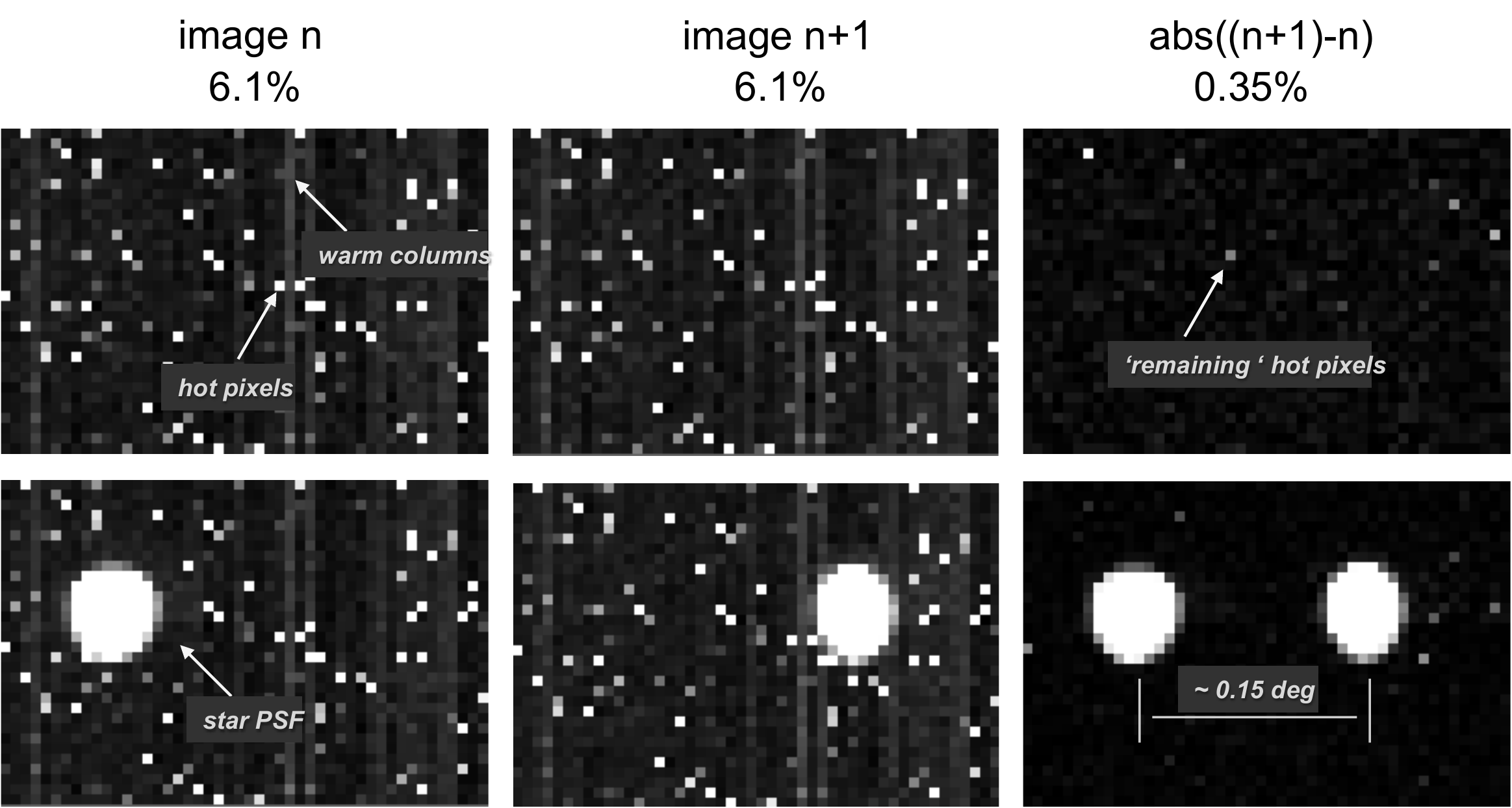}
     \caption{Illustration of the chopping procedure. Top: empty {rasters} ({24 $\times$ 36} pixel subsets of a frame). The left and middle raster was off-set horizontally by about 0.15$^\circ$ . Bottom: same as for top row, but with the telescope moved to a nearby star in the raster. Right column: absolute values of the raster differences. All images here were taken at about +20~$^\circ$C operating temperature. The values on top of the upper row are the \%  of pixels which reveal a dark current higher than 100\,ADU, compared to the median background of all pixels in the respective raster.
     }
     \label{fig5}
\end{figure}
%MDPI: is the color necessary? if not, please remove it. --which colors?

As is described in \cite{po1,po2}, a very efficient technique to overcome the mentioned detector flaws and to  significantly improve the accuracy of the CCD photometry is the ``chopping'' technique, which was introduced to the observing procedure in November 2014 and installed in February 2015 as the default observing mode for all satellites. This mode replaced the previously used ``stare'' mode. In the chopping mode a satellite is shifted between exposures back and forth, so that for every second raster-image the star is positioned in the other part of the raster (Figure\,\ref{fig5}). 
Finally, the difference of two subsequent rasters contains essentially only information relating to the stellar brightness and all local background features are close to being eliminated.

Data reduction of all \br\ photometry is the responsibility of the Data Reduction and Quality Control (DRQC) team (see Section\,\ref{orgop}). The data corrected for, for example,  the flux values with the CCD temperature and x and y pointing positions on the CCD, are archived and forwarded to the Principal Investigators (PI) for further de-correlation. De-correlation methods have been developed by Pigulski and documented as  the ``\br\ Cookbook'', which can be accessed together with the {software code at} \url{https://www.pta.edu.pl/pliki/proc/vol8/v8p175.pdf}. Examples of \br\ photometry are presented in Figures\,\ref{fig4}, 13--16, 18, 21 and 23.
%MDPI: Please add accessed date for url. Format: URL (accessed on day month year). same below --???? unclear 

\subsection{Organisation and Operation}      \label{orgop}

Organisation and operation of \bc\ relies on six interacting teams (Figure\,\ref{fig6}), which are:

\begin{figure}[H]
%\center
     \includegraphics[width=8.5cm]{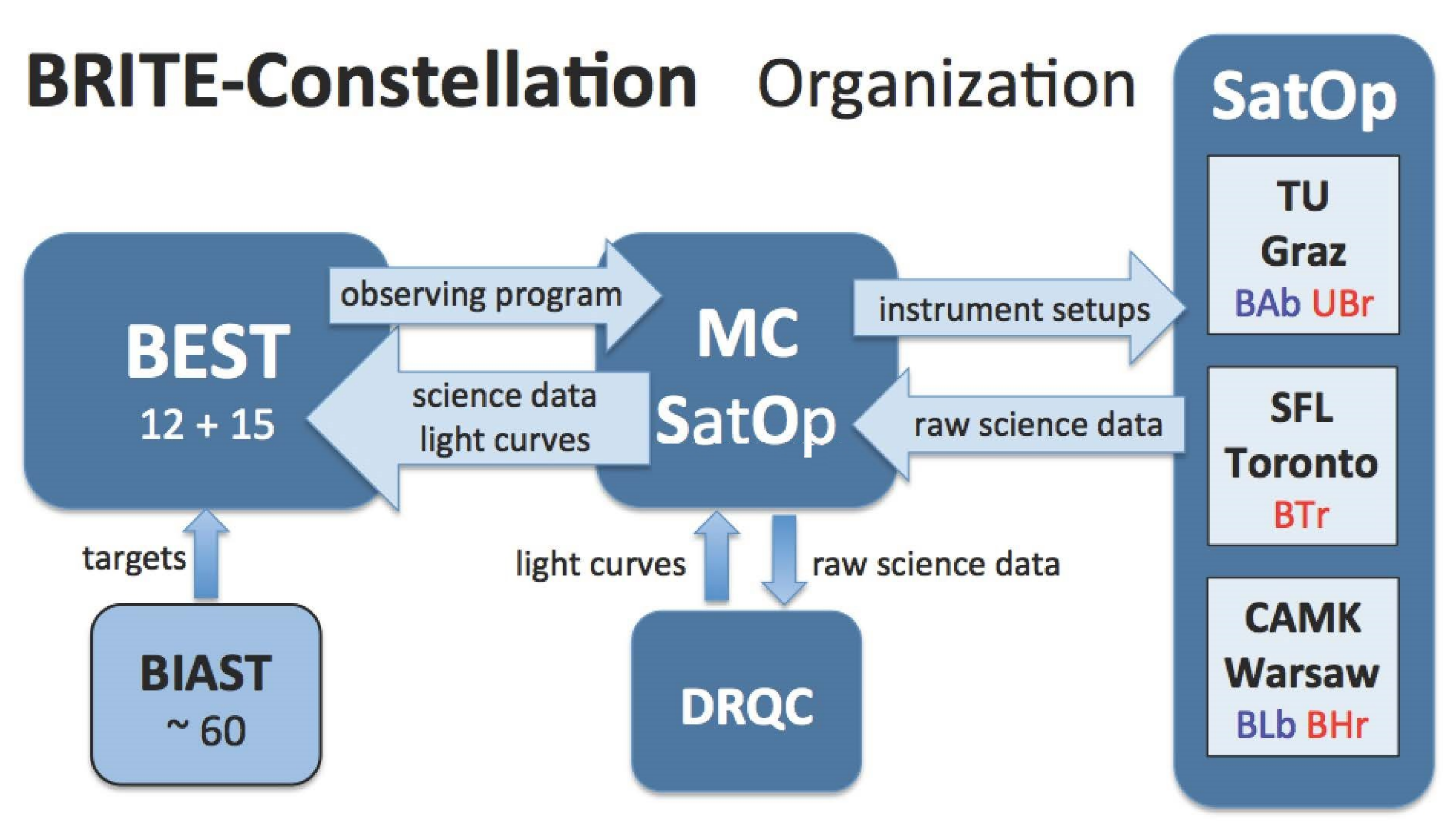}
     \vspace{-3mm}
     \caption{Organisation structure of \bc . BEST: \br\ Executive Science Team, MC: Mission Control, SatOp: Satellite Operation, DRQC: Data Reduction \& Quality Control, BIAST: \br\ International  Advisory Science Team}
     \label{fig6}
\end{figure}

\begin{itemize}
\item BEST (\br\ Executive Science Team) is the ruling body of \bc . It consists of two voting members per satellite, nominated by the three member countries (Austria, Canada and Poland) which funded the \br\ satellites.~BEST elects additional non-voting experts, presently 15. BEST releases 6 to 12 months before a new observing campaigns starts a \br\ Observing Plan (BOP), which typically covers 12 to 14 months of operation. The BOP defines which satellite is assigned to which field and for how long (Figures\,\ref{fig7} and \ref{fig8}). The rather long lead-time allows the PI's to organise supplementary observations from the ground or from space. 
\item MC (Mission Control) team is headed by Rainer Kuschnig (IKS, TU-Graz, formerly IfA Uni-Vienna) and is responsible for the execution of BOP by providing satellite orientation and instrument setup data. To ensure a maximum efficiency of \bc , a frequent quality control of all data generated with all active satellites is another core activity of MC. Such tests are applied at least twice a week and reported to BEST every second week.~In case of  problems, MC interacts directly with the corresponding satellite operator in charge.\\ A very short turn-around time between data check and satellite operation is possible, because \bc\  observes ``only'' up to 60 stars during a campaign and basically a single person inspects the data nearly in real time. The obvious benefit is a fast response to unexpected stellar variability. The best and most 
outstanding example is the serendipitous data collection from Nova Carinae 2018. Almost instantly it was apparent that \bc\ had caught the nova days before it was  discovered visually. Hence, this early volatile phase could be covered by \bc\ in an unprecedented manner, as is explained in Section\,\ref{nova}. 
\item SatOp (Satellite Operation) teams are other key elements of the mission. Satellite operators are in charge of controlling the national spacecraft via the ground stations, of which one is in Austria at TU-Graz, one in Canada at SFL-Toronto and a third one in Poland at CAMK-Warsaw. However, in case of emergency, communication is possible from each of the ground stations to any satellite to ensure uninterrupted satellite control and data management. This was, and still is, usually required during harsh weather conditions at particular ground stations or  during maintenance periods.
\item DRQC (Data Reduction and Quality Control) is another core element of the mission. The data received from each \br\ satellite on a daily basis is delivered by SatOp to MC for a preliminary quality check. Once a campaign on a given field is finished, all raw data are ASCII formatted with a FITS-like header and made available to DRQC, which generates pipeline-reduced data files (supervised by Adam Popowicz, Silesian University of Technology, Gliwice) \cite{po1}, and performs quality control (supervised by Bert Pablo, AAVSO). The original data, the raw science data (ASCII) files and the time series datasets are then submitted to the BRITE Data Archive (maintained by Andrzej Pigulski, University of Wroclaw). Most of the archive can be accessed publicly, but some data are still protected for a limited time for the corresponding PIs. The \br\ Public Data Archive can be {found at} \url{https://brite.camk.edu.pl/pub/index.html}. 
\item BIAST (\br\ International Advisory Science Team) is an informal group of presently 60 scientists, who have already successfully proposed relevant observations and/or are planning this in the future. Hence, BIAST members have expertise in \br\ data, have published the results and can advise BEST in optimising the observing program. 
\item GBOT (Ground-Based Observing Team), which is headed by Konstanze Zwintz \mbox{(U. Innsbruck)}, provides a platform for \br\ scientists and observers worldwide to support collaboration and to maximize the scientific output of \bc . 
\end{itemize}

\begin{figure}[H]
%\center
     \includegraphics[width=9cm]{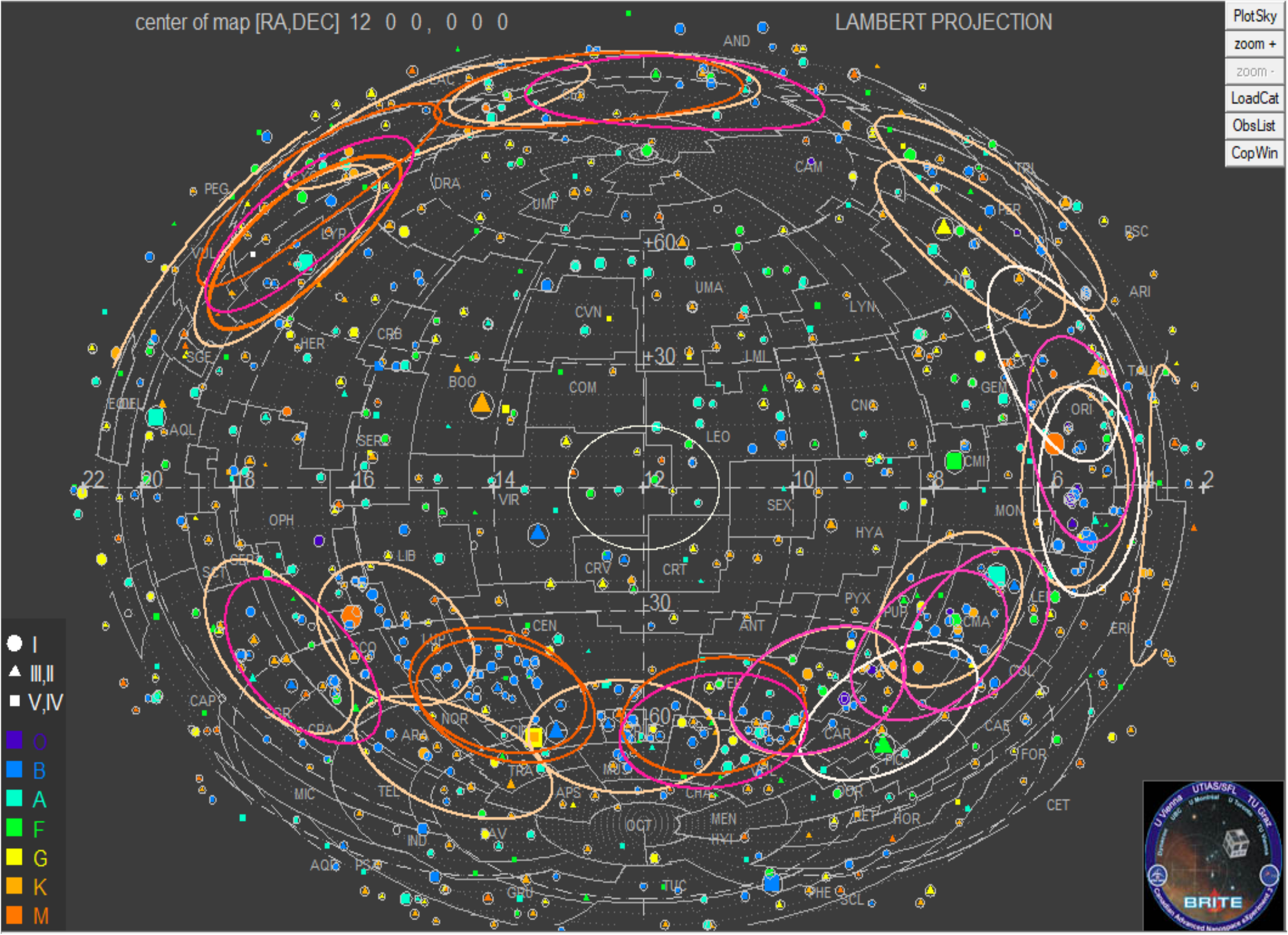}
     \caption{Sky map highlighting the fields observed thus far by at least one \br\  satellite. }
     \label{fig7}
\end{figure}
\vspace{-6pt} 

\begin{figure}[H]
\includegraphics[width=9cm, angle=-90]{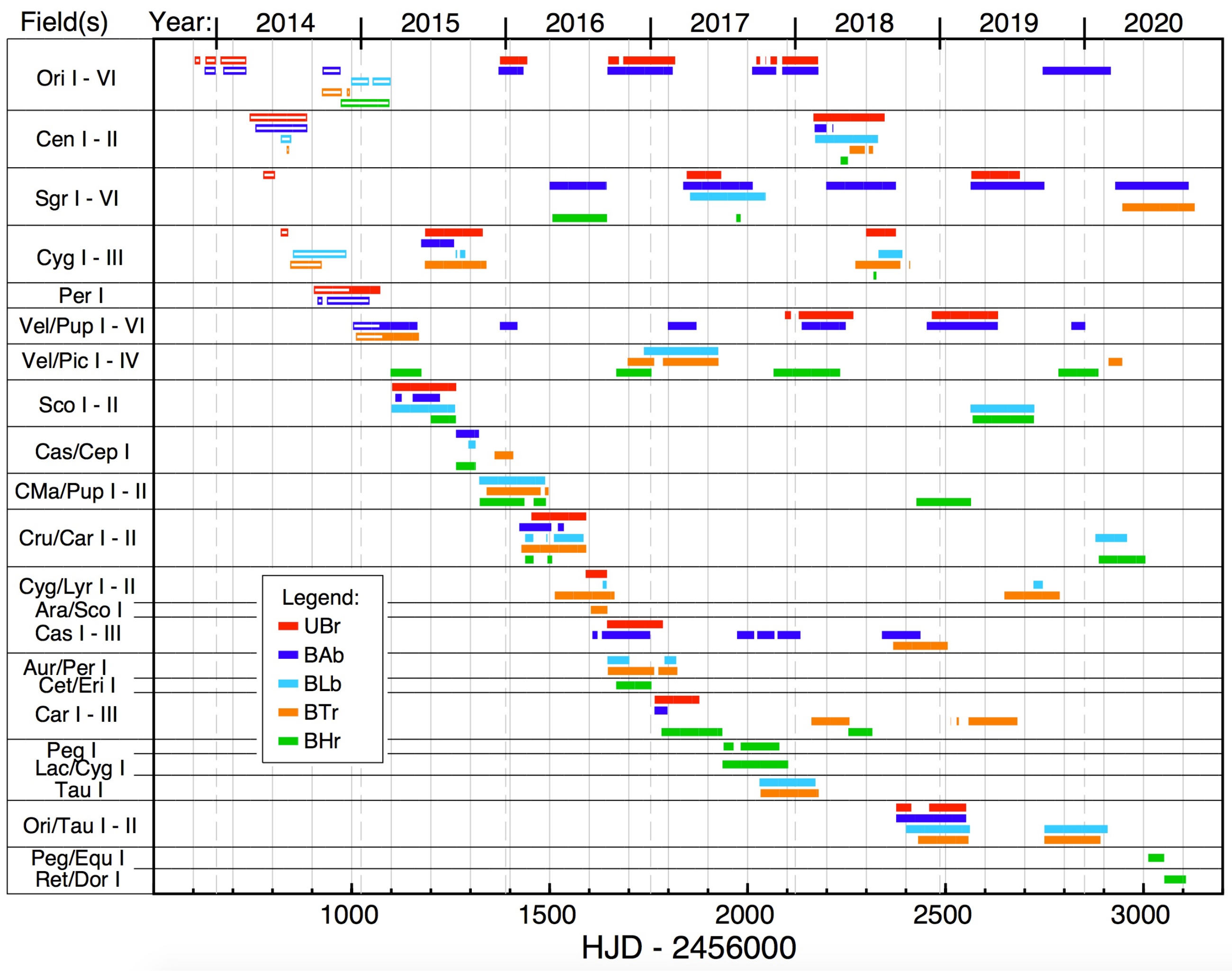}
     \caption{Temporal distribution of the observations of all five active \br\ satellites until the end of 2020. The data obtained in the stare and chopping observing modes are shown with unfilled and filled bars, respectively.}
     \label{fig8}
\end{figure}

\subsection{Present Status}  

The various star fields observed with various \br\ satellites since launch and until 2020 are presented in Figure\,\ref{fig7}. Which satellite observed a field in which period, either in chopping or stare mode, is indicated in Figure\,\ref{fig8}.

As of March 2021, 705 individual stars have been observed so far, often contemporaneously in two colours, and almost 6 million image-rasters of target stars have been produced. 
Most observations occurred in fields close to the Galactic plane, where the density of very bright stars in the FOV is high, allowing a proper choice of guide stars by the much less sensitive guiding telescope (Figure\,\ref{fig7}). Also many of the primary targets listed in the early \br\ proposals were located in this area, for example, the bright OB, B and Be stars in Orion, Carina, Centaurus or Sagittarius. 
%Presently, the 60th observing campaign is ongoing and more than 6 more are listed in the topical \br\ Observing Plan (BOP).

The observing strategy of BEST during the past eight years focused not only on stars of primary interest to the \br -community, like 6-month campaigns on hot, massive and intrinsically bright stars, but also to re-observe high profile targets essentially every possible season. The best examples are the brightest stars in the Orion field, which have been selected for the first campaign starting in December 2013 and which are currently being observed for the eighth time (Figure\,\ref{fig8}). These datasets are certainly jewels of the \bc\ legacy program. 

Even though the early \br\ science program focused on O to B (including Be) type stars, it also includes now objects beyond this range in the HR diagram (Figure 12), which is indebted to wide field photometry, reaching by default many stars and of different type. For example, cool red-giants have been observed, although not originally considered a priority, but the first data analysis led already to a relevant publication. An excellent example is $\beta$\,Pictoris (Section\,\ref{bpic}). Finally it should be mentioned that \te\ obtained data for stars which \br\ satellites observed simultaneously. An example was already given in Section\,\ref{flash} with $\zeta$\,Puppis.

For all \br\ satellites the nominal lifetime was two years. Hence, the still active satellites exceed this limit more than three times, which illustrates the high engineering quality. Nevertheless, \bc\ encountered technical problems described in the following.

The photometric accuracy is limited primarily due to stabilisation problems of \br\ satellites, but also by problems related to increasing CCD defects (\citet{po4,po5}). The development, e.g., of the normalised detector dark current with time is presented in Figure\,\ref{figdarkc}. Obviously, satellites with either a tungsten or a light weight borotron shield suffer significantly lower thermal noise increase compared to unshielded CCDs. Moreover, the sensors probably received different radiation doses during launch, as is indicated in Figure\,\ref{figdarkc}  by the onset of the linear approximations.
% \vspace{-10pt} 
\begin{figure}[H]
%\center
     \includegraphics[width=8.5cm]{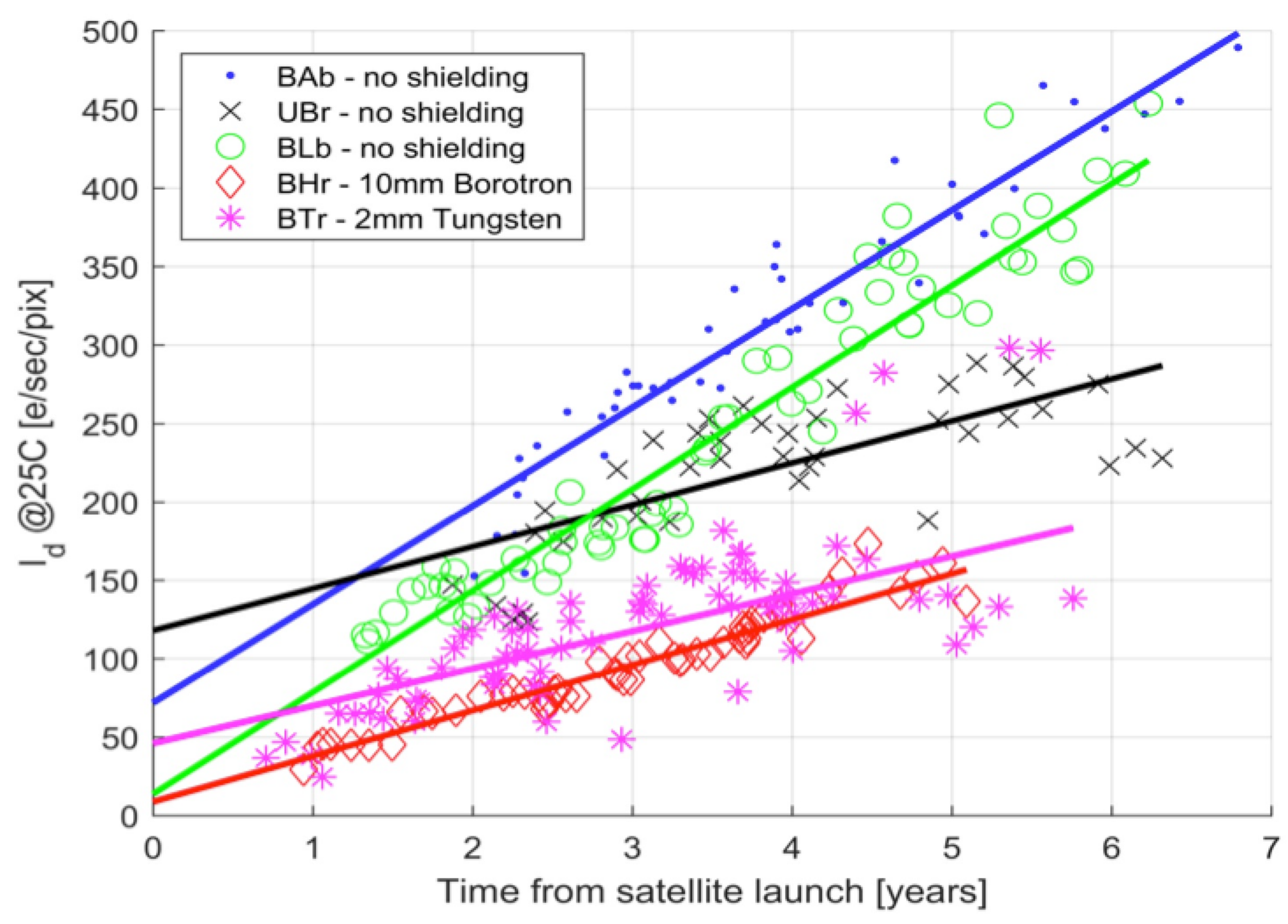}
     \caption{Temporal development of the CCD dark 
     currents.
     }
     \label{figdarkc}
\end{figure}

\clearpage
The status of individual \br\ satellites can be summarized as: 

\begin{itemize}
\item	BRITE-Toronto (BTr), is in good condition and produces among the best data, despite a significant amount of radiation damage. Primary target stars can be placed on the CCD where the background is least noisy. 
\item	BRITE-Heweliusz (BHr), is working very well in general; some observing fields seem to cause problems for the pointing system, but usually alternative orientations of the field (different guide stars) can be chosen. It also has the least amount of radiation damage due to a better shielding of the CCD. 
\item	BRITE-AUSTRIA (BAb) produces scientifically relevant data, even after more than eight years in orbit and an enhanced radiation environment. To obtain the best photometric consistency over the lifetime of \bc , this satellite has been assigned to observe every year essentially the same set of fields in Orion \mbox{and~Sagittarius.} 
\item	UniBRITE  (UBr), was working well until June 2019, despite its high grade of radiation damage. However, it failed after that date and a failure analysis led by SFL and conducted by IKS TU-Graz concluded that one of the three reaction wheels seems to be damaged and cannot be used for stabilising the spacecraft. A repair concept is being developed. 
\item	BRITE-Lem (BLb), worked well until April 2020 when it consistently failed to get into fine pointing. This is very likely due to a damaged reaction wheel. However  more tests are still to be conducted to come to a firm decision.
\end{itemize}

In conclusion, presently three of the five functioning \bc\ satellites are still operational: BHr and BTr are producing very good data and BAb still useful photometry. BEST  expects to continue the mission until at least in 2022, depending on unpredictable technical failures, for example, of the reaction wheels. Attempts to recover the other two \br s will continue.

%============================================================================
\section{Key Results of the Mission and Scientific Highlights}  \label{Highlights}
%============================================================================

Since its launch \bc\ has obtained measurements for 705 individual targets in 60 currently completed fields (Figure\,\ref{fig7}) of which many overlap. A large fraction of the targets was observed in more than one field which yields total time bases of up to eight years for several stars (Figure\,\ref{fig8}). As of March 2021, 11.5\,\% of all targets observed by \bc\ are included in one or more peer-reviewed publications.
%, which compares well with \mo , \co , \ke\ and \ke-K2. 
\br\ data of many other targets are still being actively analyzed and will be the topics of additional future papers.
In the following, selected research highlights based on \bc\ data are presented, mostly sorted from most massive to least massive stars. The individual objects are also indicated in Figure\,\ref{fig9}. 

\begin{figure}[H]
%\center
     \includegraphics[width=7.3cm]{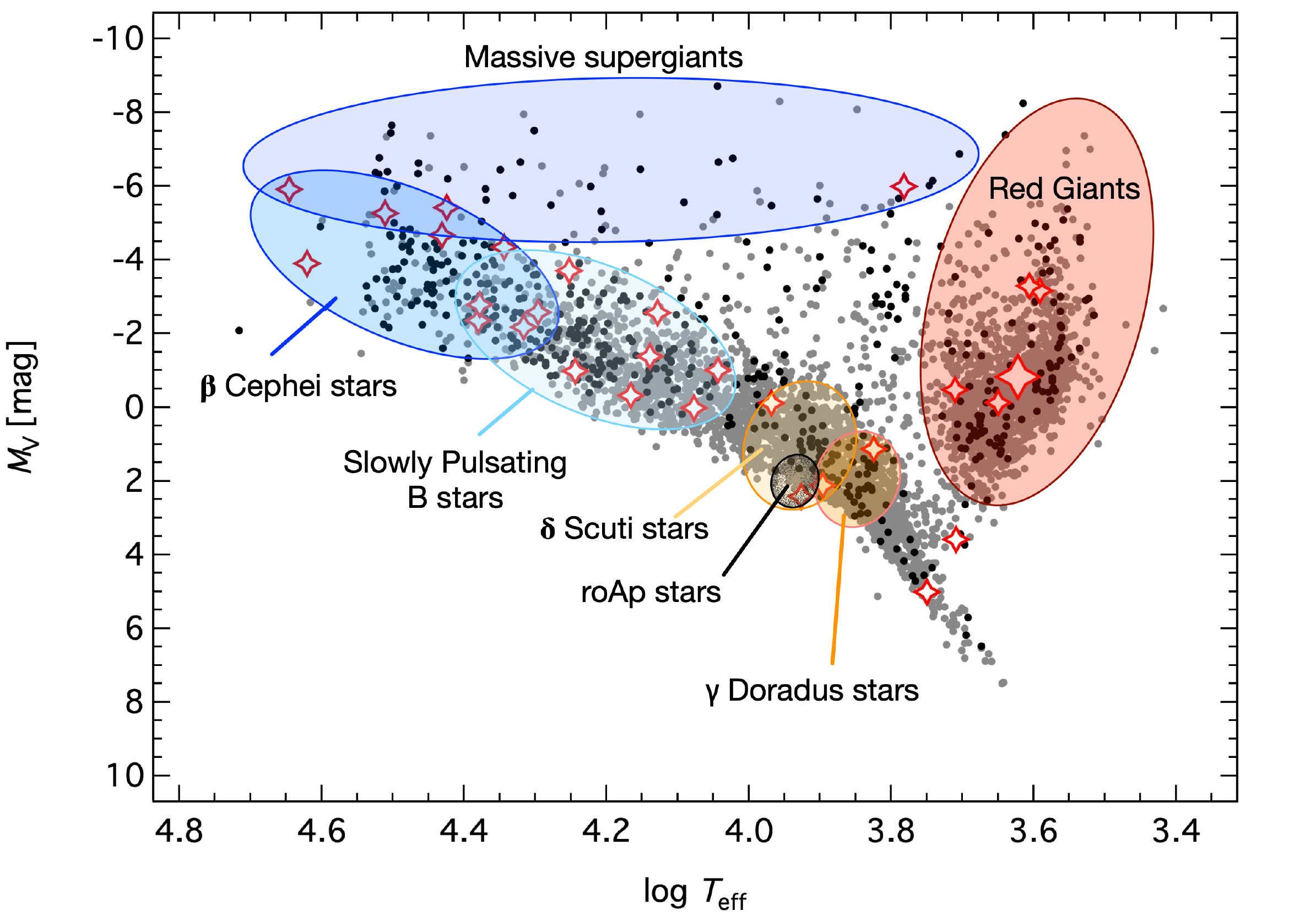}
     \caption{{HR diagram} of the stars brighter than 6th~mag in $V$ (grey dots). Stars for which BRITE photometry was collected are shown by black dots. The objects discussed in Section \ref{Highlights} are marked as open red symbols where the larger symbol stands as a representation of the 23 red giants discussed in Section \ref{redgiants}. Indicative instability domains for several types of pulsators are shown as colored ellipses. Be stars cover much of the $\beta$\,Cep and SPB domains}
     \label{fig9}
\end{figure}
%MDPI: in the image, Please change hyphen (-) to minus (−) in numbers --presently not possible to do 

\subsection{The Link between Stellar and Wind Variability in Very Massive Stars} \label{szPup}

High-precision photometry of the runaway early-O-type supergiant $\zeta$\,Puppis \linebreak (\mbox{Figures\,\ref{figzpup}, \ref{figTEBR} and \ref{fig15}}) revealed that a previously-proposed rotation period of 5.1\,d is incorrect and the period actually is 1.78\,d, which agrees much better with a model for the rotational evolution. 
Figure\,\ref{fig15} also indicates that the large, real scatter beyond the 1.78\,d modulation, is probably due to stochastically varying short-lived bright regions in the photosphere arising in a subsurface convection zone, which led to clumps in the wind. An alternative supposition is that the stochastic variability arises from gravity waves at the internal radiative/convection border. This is supported by hydrodynamic simulations showing  gravity waves  causing stochastic variability in the photospheres of main sequence OB stars (\citet{bow11}).
\begin{figure}[H]
%\center
     \includegraphics[width=70mm, angle=-90]{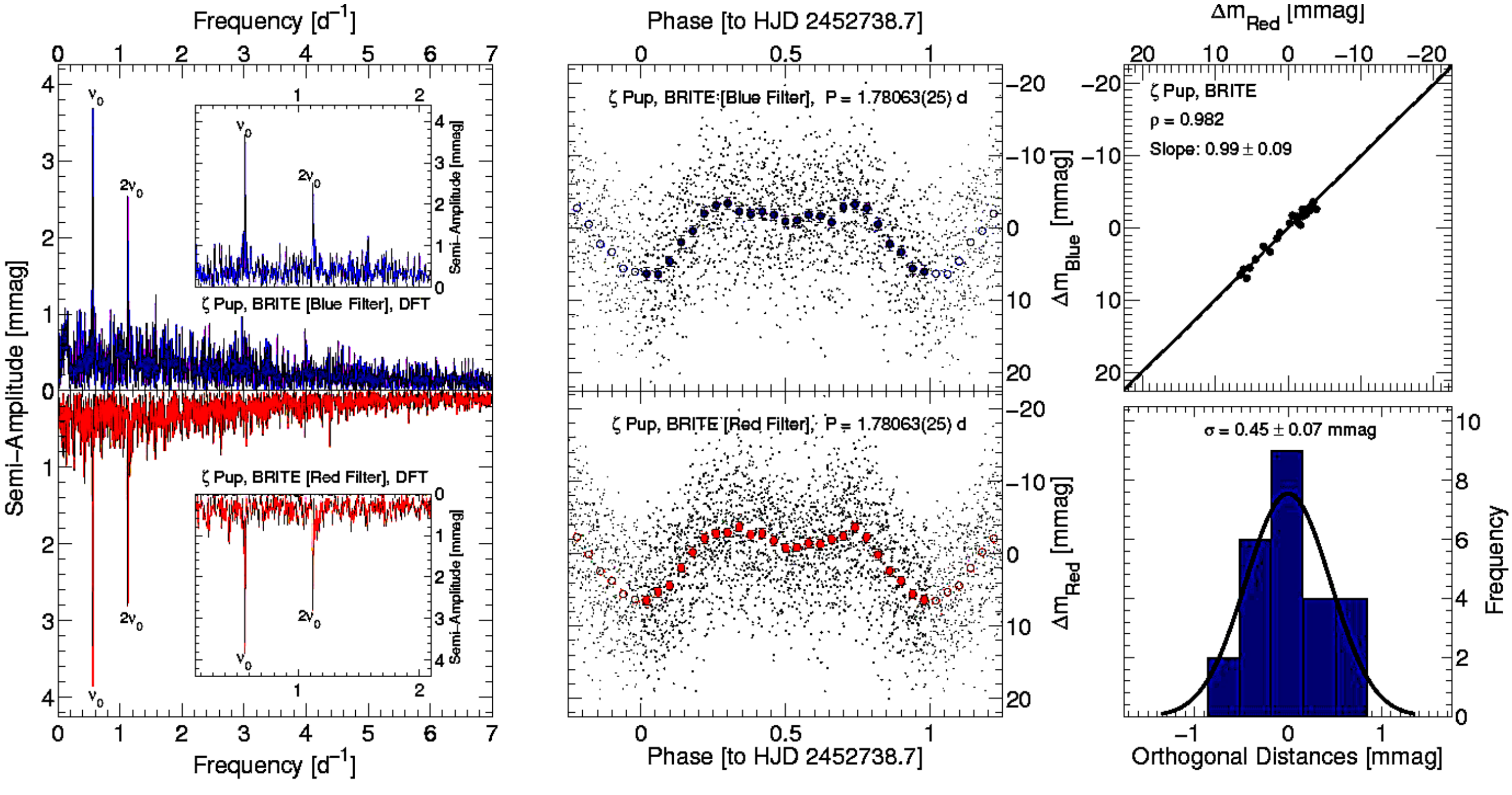}

     \caption{\textbf{Left}: Fourier transform of the red and blue 2014/15 \br\ photometry of $\zeta$\,Puppis (see Figure\,\ref{fig8} for VelPup I-VI runs).~\textbf{Middle}: Corresponding rotation light curve (P = 1.78\,d). Coloured points are 0.04 phase bins with 1\,$\sigma$ error. 
     Note the large, real scatter, which could be due to stochastically varying short-lived bright regions in the photosphere which lead to clumps in the wind.
     \textbf{Right}: Comparison diagram for the phased blue and red light curves (\textbf{top}), and  distribution of orthogonal distances with a Gaussian fit (\textbf{bottom}). (Figure\,5 of \citet{rami1}). }
     \label{fig15}
\end{figure}

The top diagram in the right column of Figure\,\ref{fig15} shows that both kinds of bright spots show the same variability amplitude in the \br\ blue and red filters, implying  insensitivity to the expected hotter nature of the spots compared to their adjacent areas in the stellar photosphere. The reason for this is that the Rayleigh-Jeans tails of the stellar emission spectrum  are sampled at significantly longer wavelengths, compared to the UV maximum peak. The bottom diagram in the right column is consistent with the photometric precision of the data.

The findings yielded by the 2014/2015 observing campaign on $\zeta$\,Puppis may be an important resolution of a long-standing puzzle indicating subsurface convection as the main source of the two types of wind variability (quasi-periodic co-rotating interaction regions-CIRs-and stochastic clumps), which previously was not considered possible in such hot stars.

After this study of $\zeta$\,Puppis, parallel observations were obtained in 2018/19 using \br\ in the optical and  \cha\ in X-rays (\citet{nich1}). Both satellites confirm a 1.78\,d period (Figure\,\ref{fig13}), which is thought to be the result of bright photospheric spots driving CIRs in the stellar wind, with the X-rays arising somewhat further out in the wind, where the CIR shock is strongest. Alternatively, as noted above, the stochastic component of variability could arise from gravity waves arriving from a much deeper zone.

\begin{figure}[H]
%\center
     \includegraphics[width=11cm]{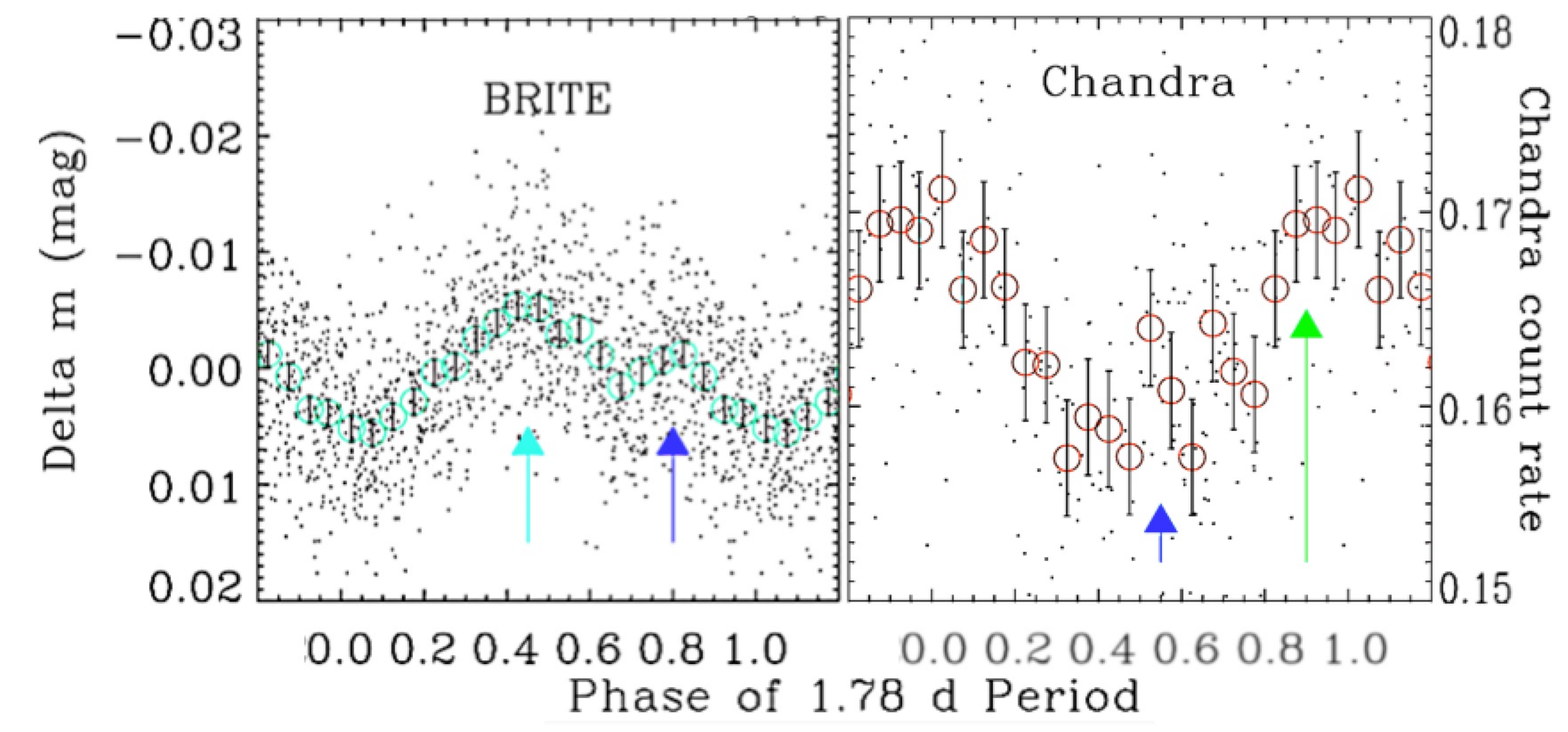}
     \caption{$\zeta$\,Puppis observed in the visible with \br\ and in X-rays with \cha , folded with the  period of 1.78\,d. The multi-wavelength light curve behaviour presumably illustrates the effects of Corotating Interaction Regions. The cyan arrows indicate the primary and the blue arrows the secondary maximum. There is a significant shift in the times of maximum due to a large delay or a smaller shift but mismatch in which is primary and secondary maximum  (Figure\,3 of~\mbox{\citet{nich1}}). }
     \label{fig13}
\end{figure}

A very recent investigation was made on about 60 bright galactic Wolf-Rayet stars using combined data sets from \mo , \br\  and \te\ by Lenoir-Craig et~al. (submitted to {ApJ} and in \cite{len1}).  Fourier analysis of the light curves reveals an important trend of enhanced stochastic variability at low frequencies ($\lesssim$1\,\cd ) with the spectrally-modelled hydrostatic-core temperatures (T$^\ast$), much like a preceding ground-based spectral variability study by \citet{che1, che2}. Both studies support the idea that the stochastic variability seen in all WR stars arises in clump formation and propagation in their strong winds, such that, surprisingly, hotter WR stars with faster winds show less variability and hence less clumping. This can be explained by the triggering of the clumps in subsurface convective zones that are deeper and stronger in cooler WR stars. This may or may not conflict with the heretofore theory of clump formation by wind instabilities, which are expected to be stronger in hotter, faster WR winds.

{ Other targets with similar science relevance are WR\,40 (WN8h; \citet{ta5}), V973\,Sco (O8Iaf; \citet{ta6}) and $\gamma ^2$\,Vel (WC8+O7.5III-V; \citet{no5}). They were among those prominently observed by \bc\ during several runs and helped to investigate the dynamics of winds and their relation to variations occurring at the stellar (hydrostatic) surface.}

\subsection{The Heartbeat of Stars: $\iota$\,Orionis and $\epsilon$\,Lupi}

Heartbeat stars are a class of eccentric binaries which are characterized by tidally excited oscillations (TEO) with distinct amplitude changes at periastron. They are uniquely interesting for the study of massive stars, because they allow for full binary solutions without eclipses and provide access to asteroseismology of objects where pulsation is rare. 
Using \bc , the well-studied binary system $\iota$\,Ori (O9III+B1 III/IV) was the first massive star ever in which TEOs were discovered, and which opened a whole new avenue to studying massive star interiors (\citet{pab2}). 
The data in Figure~\ref{fig20} are phased to periastron (phase = 1.0, with P = 29.13376\,d) and binned to 0.0025 in phase.  

Another unique heartbeat star discovered with \bc\ was $\epsilon$\,Lupi. \textls[-15]{This system is the only known doubly magnetic massive binary (\citet{shu1}). \mbox{\citet{pab3}}} were able to determine masses and radii despite an orbital inclination of $\approx$20$^\circ$. This allows one to explore the interesting interplay between magnetism and tidal effects in the evolution of such a system. 
%Using \bc , this was the first massive star in which TEOs were discovered, which opened a whole new avenue to studying massive star interiors (\citet{pab2}). 

\begin{figure}[H]
%\center
     \includegraphics[width=10cm]{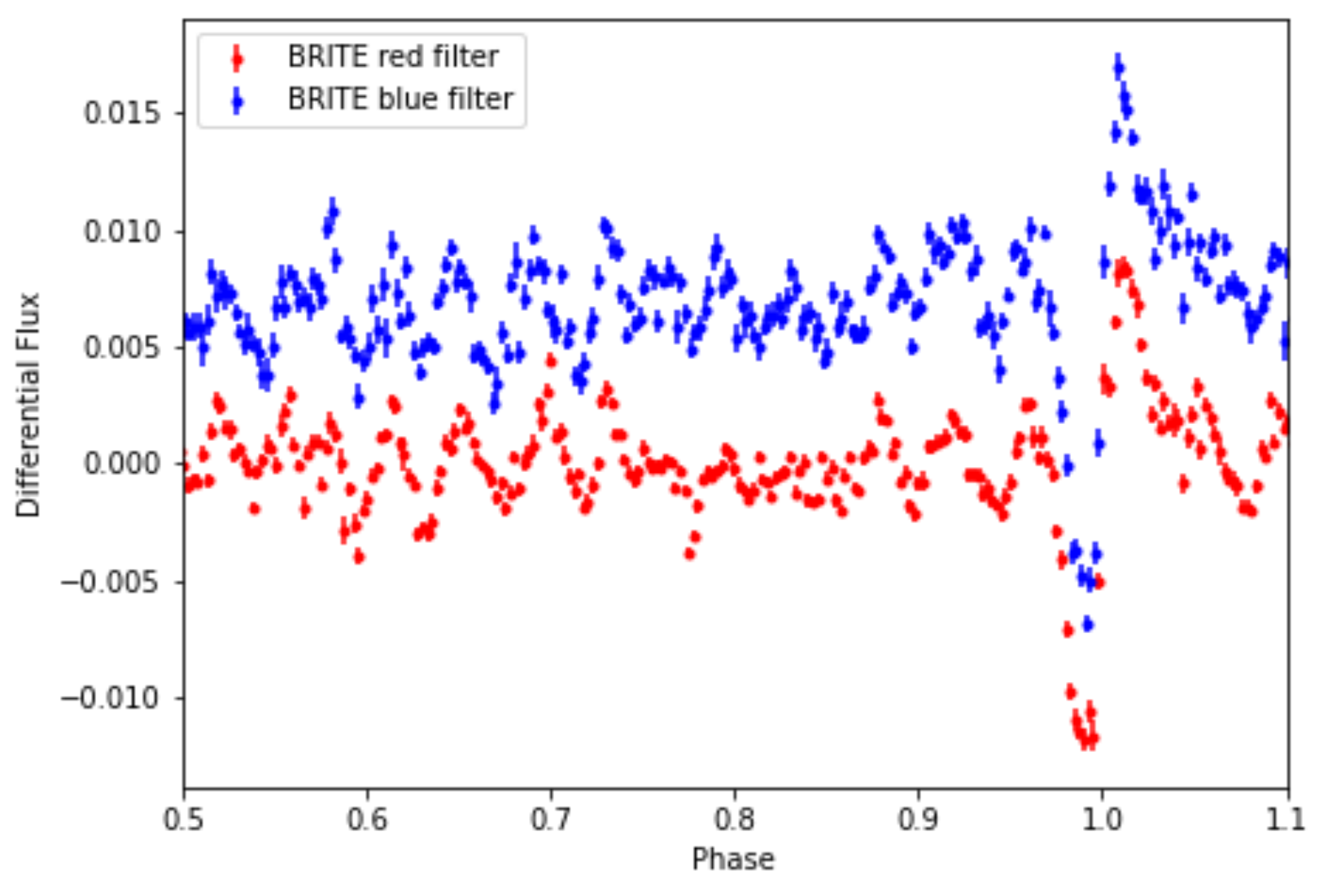}
     \caption{Binned and phased data of $\iota$\,Ori, obtained with \bc , covering two years. The data clearly show tidally excited oscillations, most prominently from 0.5--0.8 in phase, as well as a strong heartbeat signal (0.95--1.05 in phase). The blue points are shifted by a constant flux for clarity.}
     \label{fig20}
\end{figure}

The value of \br\ heartbeat stars also extends to the upper reaches of the HR diagram with the enigmatic and highly eccentric binary system $\eta$\,Car, although the length of the period combined with mass loss have made it difficult to characterize any heartbeat signal at periastron. Using two separate \br\ observations, \citet{ric5} were able to confirm oscillation frequencies, which appear to be stable over the past four decades (\citet{vge5}, \citet{ste5}).
These frequencies share many similarities with TEOs, though this identification will need more data to confirm.

\subsection{The Riddle of Betelgeuse}

The red supergiant Betelgeuse is not only one of the biggest stars in the sky, but also one of the most puzzling. Long-term photometry and radial velocity studies reveal semi-regular stellar pulsation periods of 420\,d, and possibly superposed by a cycle of 8.7\,years (\citet{gol1, dup1, smi1}). In comparison, \citet{kis1} report $388\pm 30$\,days as a pulsation period and a $5.6\pm 1.1$\,years cycle, using AAVSO-V data obtained almost during an entire century (1918--2006).

Curiously, its high apparent brightness makes Betelgeuse a difficult target for ground-based photometry, as big telescopes suffer from over-exposure. This gap is now filled with 
high-quality \br\ photometry (Figure\,\ref{figAORI}), %The AAVSO data base was most helpful to bring \br\ data to a common photometric system.
augmented with spectroscopic data obtained during more than 10 years at the STELLA robotic observatory, which is one of the biggest fully robotic telescopes worldwide (\citet{str9}). Only automated observing procedures allow scheduling of almost daily visits of the same star, each lasting no longer than 5 min and stretching over more than a decade. More than 2000 individual, high-resolution spectra have been collected and automatically reduced.

As Figure\,\ref{figAORI} illustrates,  the radial velocity variations follow in general closely the photometry, suggesting a physical link between photometric and radial velocity variations. 
A splitting of the main pulsation period into P\,$\approx 420$\,d and P\,$\approx 335$\,d, as seen in the RV-data, is proven beyond doubt by BRITE photometry, which provides the required cadence and precision. This is work in progress-adding another season will hopefully put these findings on solid grounds.

\begin{figure}[H]

     \includegraphics[width=13cm]{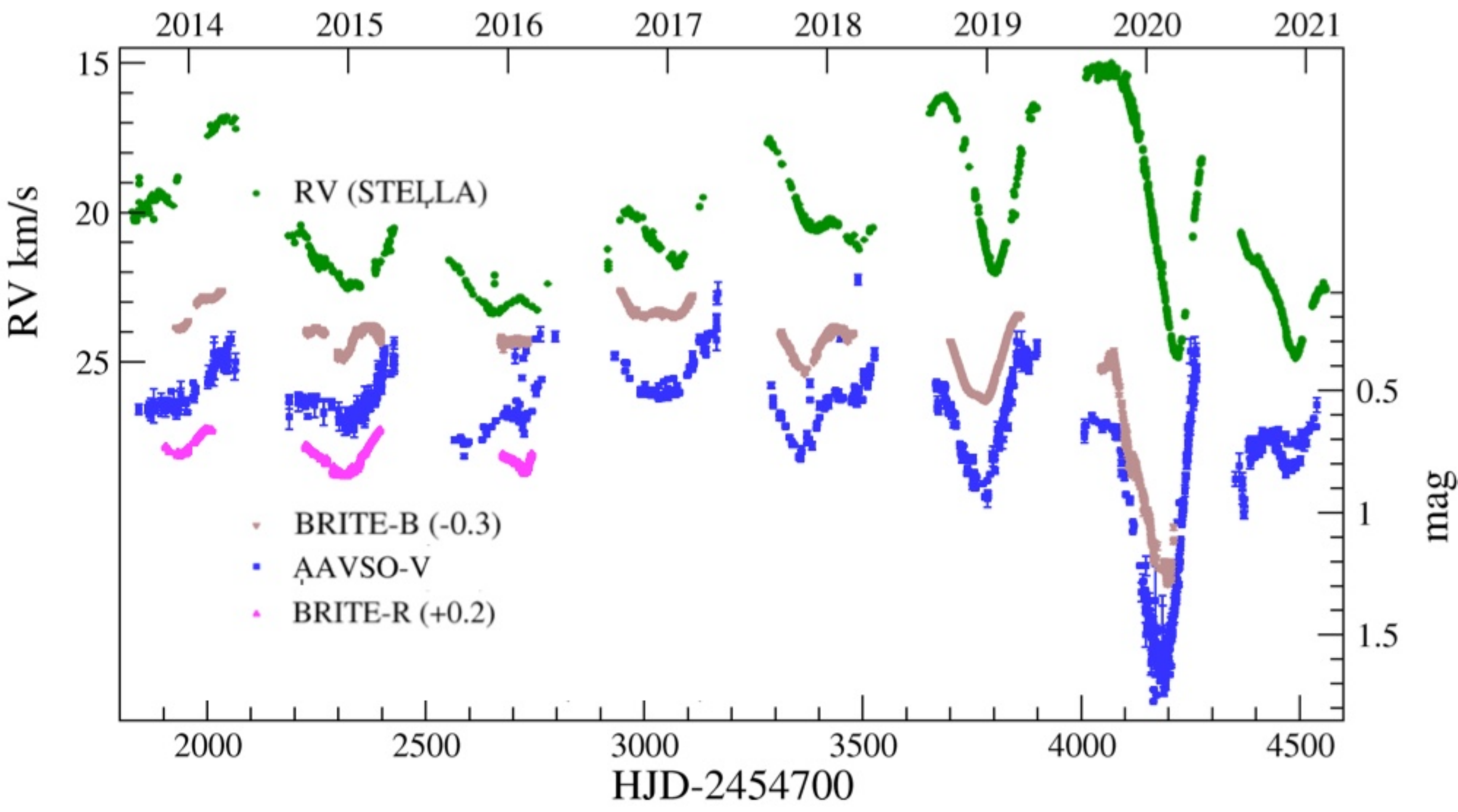}
     \caption{Comparison of light and RV variations of Betelgeuse. From top to bottom: STELLA RV data, \br -blue, AAVSO-V absolute photometry, and \br -red photometry. Error bars on \br\ magnitudes and on STELLA RV are too small to be visible.
     %Comparison of \br\ B \& R magnitudes to radial velocities, measured with the robotic telescope STELLA. Absolute photometry from the AAVSO database is plotted in blue (between \br -B and \br -R). Error bars on \br\ photometry and STELLA RV are too small to be visible.
     }
     \label{figAORI}
\end{figure}

During the grand dimming event in the 2020/21 season, an extraordinarily excursion of the photometry to the RV-data can be seen.
The photometric amplitude by far outstretches the already high RV amplitude. An analysis of HST UV-data of this period (\citet{dup2}) hints to a big plume of dust being emitted from the surface of the star and subsequently drifting into the line of sight, thereby enhancing the photometric~minimum.

Betelgeuse is approaching its end of life as a star, commonly believed to be a supernova progenitor. Last year's dimming event sparked estimates that an explosion may be imminent within the next 
100,000 years. But  observations and models are currently not refined enough to prove whether Betelgeuse will end in a type IIb, II-L, or II-P supernova (\citet{mey1}). Hence, new observations are needed to better estimate mass and rotation rate in order to pin down Betelgeuse's future path. \bc\ will participate in these campaigns.

\subsection{Evolving Pulsation of the Slowly Rotating Magnetic $\beta$\,Cephei Pulsator $\xi^1$\,CMa} 

$\xi^1$\,CMa is a remarkable magnetic early B-type star that is distinguished in several ways: it rotates extremely slowly ($P_{\rm rot}\sim 30$~y; \citet{shu2}), it is the only magnetic B-star known to exhibit detectable H$\alpha$ emission from a dynamical magnetosphere \cite{shu2}, and its optical and X-ray magnetospheric emission are modulated according to its $\sim$0.2\,d radial pulsation period (\citet{shu2,osk1}).

Building on work by \citet{pigu2,jer2} and \citet{shu2}, \bc\ photometry (BLb, BHr, BTr) of $\xi^1$\,CMa was employed by \citet{wad1}, as one of the most recent anchor points to monitor the evolution of its pulsation period. Combining over one century of photometric and radial velocity monitoring, they concluded that the period evolution of $\xi^1$\,CMa consists of a secular period lengthening of roughly \mbox{0.3 s/century} that can be satisfactorily understood as a consequence of expansion due to stellar evolution. An additional period evolution---more rapid and of lower amplitude---remains unexplained, and the authors speculate that it may be a consequence of rotational modulation or evolution that is restricted to relatively rapid, short-term episodes, rather than uniform long-term changes. Binarity can be ruled out, because the corresponding RV variations would have been easily detected.

\newpage
\subsection{The Triple System $\beta$ Centauri}    %%%%%%%%%%%%%%%%%%%

\textls[-5]{Massive stars, with initial masses greater than 8 $M_{\odot}$, are among the least understood, but they are extremely important, because they produce the majority of heavy elements. A fascinating BRITE-Constellation target is the triple system $\beta$\,Centauri \mbox{(Figure\,\ref{figAP})}---also named Agena--consisting of a massive binary ($\beta$\,Cen\,AaAb: B1 II and B1 III) with an eccentric orbit and a more distant and 3 mag fainter companion, also of B-type (\mbox{\citet{pigu1}}).}

\begin{figure}[H]

     \includegraphics[width=12cm]{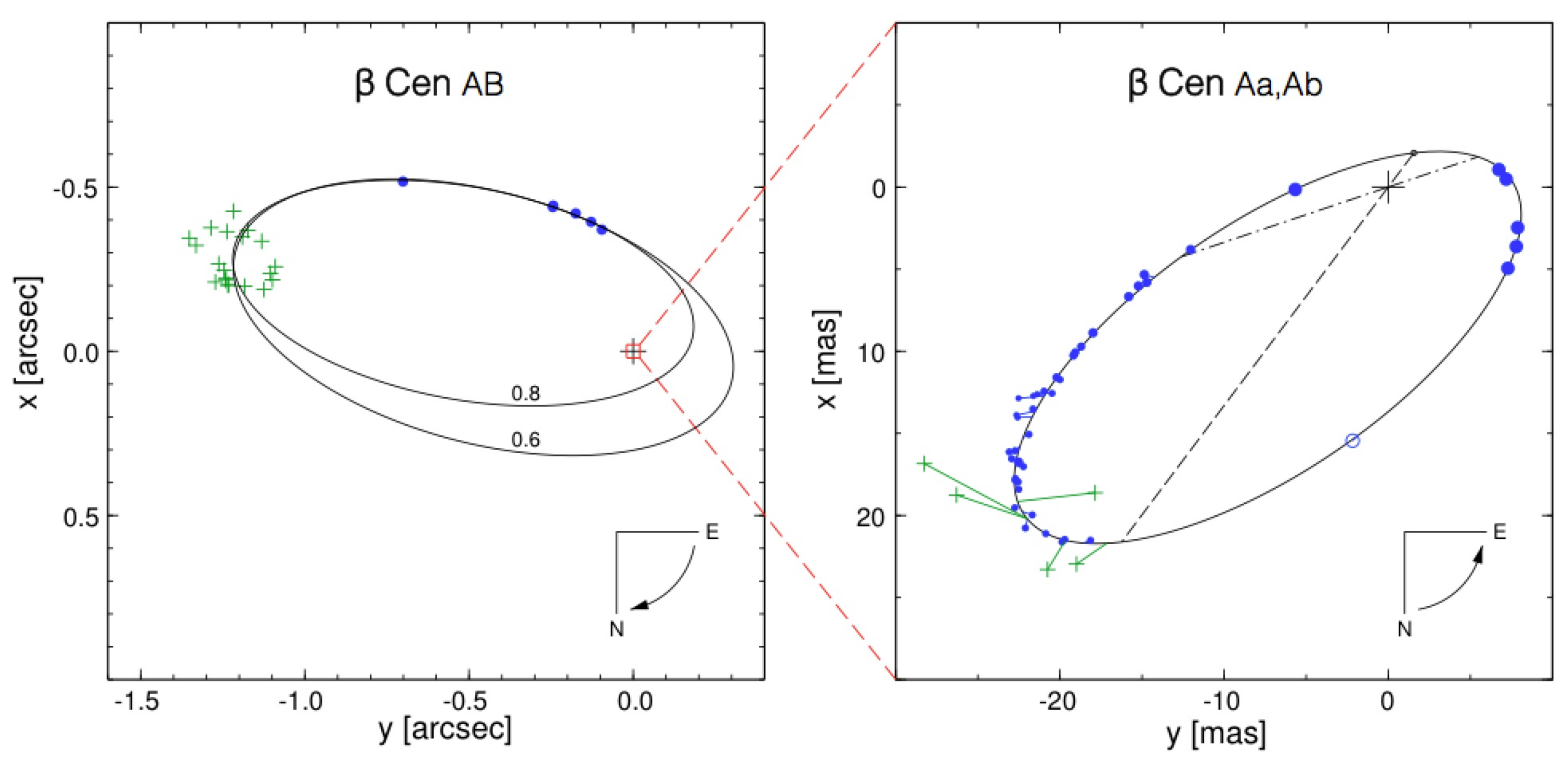}
     \caption{{The triple} system $\beta$\,Centauri, with two example orbits (excentricity of 0.6 and 0.8). Adapted from Figures 1 and 2 of \citet{pigu1}.}
     \label{figAP}
\end{figure}
%MDPI: in the image, Please change hyphen (-) to minus (−) in numbers --presently not possible to do 

$\beta$\,Cen\,B was discovered in 2011 as a magnetic star (\citet{ale1}).
\mbox{With 17 detected} p and g modes, the close massive binary system becomes one of about a dozen known hybrid $\beta$ Cep/SPB stars with such a rich frequency spectrum. Furthermore, its binarity provides a very precise determination of the masses of the components, but complicates seismic modeling, because the modes need to be safely assigned to one of the components, which---in addition---are relatively fast rotating.

The case of $\beta$\,Cen illustrates the potential of \bc\ data for the detection of rich-frequency spectra of small-amplitude modes in pulsating stars.

 \subsection{Long-Period Oscillations in the $\beta$ Cephei Pulsators $\nu$ Eridani and $\theta$ Ophiuchi}   
  
  \textls[-20]{Thanks to the long-term stability of \br , \citet{han1}) detected \mbox{several previously}} unknown long-period signals corresponding to gravity-mode oscillations of the $\beta$ Cephei pulsator $\nu$ Eridani.
\citet{das1, das2} demonstrated that present standard pulsation models cannot reproduce the observed frequency range of g-mode pulsations, which is likely due to shortcomings in the underlying stellar physics data, in particular of opacities.

Upon the detection of a large number of g-mode pulsations in the BRITE data of another $\beta$ Cephei star, $\theta$ Ophiuchi, \citet{wal19} arrived at an identical conclusion (with the caveat that a B5 companion star could be responsible for the g modes), namely that opacities need to be increased between 30\% and 145\% (!!) in the range log \mbox{$T = 5.06$--$5.47$} to reproduce the observations. Obviously, the use of correct opacity data is important for modelling of all kinds of stars. Hence, the implied revision of these data impacts stellar physics in general.

\subsection{The Ellipsoidal SPB Variable $\pi^5$ Orionis}

 BRITE observations of the ellipsoidal variable $\pi^5$\,Orionis (\citet{jer1}) revealed that the primary star belongs to the class of Slowly Pulsating B (SPB) stars. Within the modes of pulsation, there is a re-occurring splitting of twice the orbital frequency. This is interpreted as perturbation of nonradial pulsation modes by the equilibrium tide exerted by the companion. The behaviour of the two tidally disturbed pulsation modes is largely consistent with axisymmetric dipole modes ($l=1, m=0$). These findings have two important and interesting consequences:  
 
 \begin{itemize}
\item[--] $\pi^5$\,Ori is the first SPB star in which tidal perturbations have been identified and

\item[--] these perturbations facilitate the identification of nonradial pulsation modes. \\
\br\ allowed a valuable proof-of-concept of mode identification to be carried out, which opened up tidal asteroseismology of SPB stars in multiple systems.
 \end{itemize}

\subsection{Be Stars}

The \br\ database is rich in Be-star observations because there are many bright Be stars and, for B-type stars, the blue- and red-sensitive \br\ satellites achieve roughly equal S/N, unless an extreme reddening is present.  The combination of the frequency resolution and quality of \br\ observations over several seasons with the long-term behaviour documented by \sm\ has achieved qualitatively new insights into the so-called Be phenomenon.  

Two central questions about Be stars (see \citet{riv1} for a review) are: 

\begin{itemize}

\item[(i)]  How do Be stars maintain their Keplerian decretion disks where the eponymous emission lines form and that, without regular replenishment, dissipate within a year? 
\item[(ii)] How have Be stars acquired their $\gtrsim$75\% critical rotation?  
One way to explain the latter is mass transfer in a close binary.  The former primaries often appear as hot, subluminous sdO stars that are challenging to detect even in UV spectra (\mbox{\citet{wan18}}) and contribute little flux in the \br\ passbands.  However, first photometric Doppler shifts derived from \br\ and \sm\ data spanning 25 years have set an upper limit of $\sim$1\,$M_\odot$ on the mass of a putative companion of $\nu$\,Pup (\citet{baa2}). 
\end{itemize}
%Other Be stars, though, have been found by \br\ to be relatively poor clocks.

\bc\ has been instrumental in confirming earlier suggestions that Be disks are fed by discrete mass-loss outbursts driven by the superposition of several low-order non-radial pulsation modes or by recently detected stochastically-excited pulsations, transporting angular momentum from the stellar core to the surface (\citet{nei5}).
Although originally detected in H$\alpha$ line profiles of $\mu$\,Cen (\citet{riv2}), optical photometry is a better tracer of outbursts because the $V$-band flux responds sensitively to varying amounts of ejecta causing electron scattering and free-bound recombination (\citet{haub1}).  In fact, in $\mu$\,Cen, outbursts have up to 100 times higher amplitude than the underlying non-radial pulsation and can render the pulsations undetectable (\mbox{\citet{baa16}}).  In $\eta$\,Cen (\cite{baa16}), 28\,Cyg (\citet{baa1}), and 25\,Ori (\citet{baa18}), \br\ found closely spaced NRP frequencies the difference between which corresponds to the repeat frequency of the outbursts.  During an outburst, the combined amplitude of the involved non-radial pulsation modes grows nonlinearly, demonstrating that the outbursts are pulsation powered far beyond mere mode beating.  Hierarchically nested frequency groups can drive repetitive outbursts on timescales from weeks to years (Figure\,\ref{fig:25Ori}), and the frequency groups typical of Be stars can be understood as difference frequencies (g0), non-radial pulsation frequencies proper (g1), and sum/harmonic frequencies (g2) (\cite{baa18}).  So-called {\v S}tefl frequencies first found in emission lines (\citet{ste98}) probably are orbital frequencies in the innermost inhomogeneous disk (\citet{baa16}).

For shorter timescales/lower amplitudes, TESS (\citet{lab20, lab21}) has confirmed the correlation between increased non-radial pulsation amplitude and mean brightness.  Similarly tight networks of selected non-radial pulsation frequencies do not seem to be known from other stars, and the outbursts may enable Be stars to escape an angular-momentum crisis possibly caused by the contracting core (\citet{baa20}).  The detection of non-radial pulsation modes by \br\ (\citet{bor1}) and TESS (\citet{lab21}) has also terminated decades-long speculations that the best-known Be star, $\gamma$\,Cas, shows rotational variability due to a magnetic field \mbox{(\citet{smi21})}. 

\begin{figure}[H]

     \includegraphics[width=12cm]{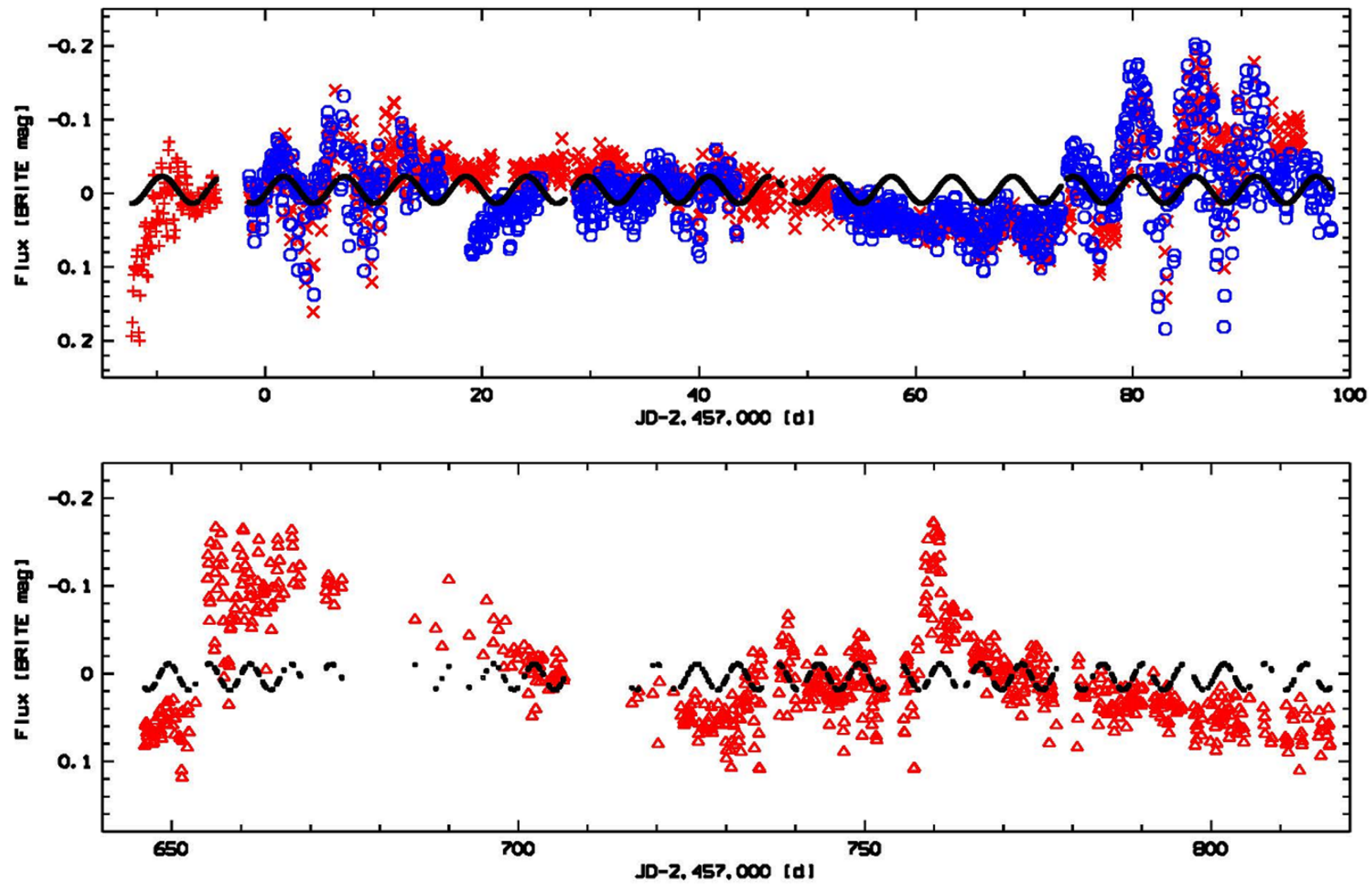}
\caption{{Blue (blue symbols)} and red (red symbols) \br\ light curves of the Be star 25\,Ori (cf., \citet{baa18}) from 2014/15 (\textbf{top}) and 2016/17 (\textbf{bottom}).  In either season, two outbursts separated by $\sim$78\,d occurred which corresponds to the difference frequency of 0.0129\,c/d between many pairs of non-radial pulsation modes.  The black curves are a sine fit to the 2014/15 light curve outside the outbursts with frequency 0.1777\,c/d which is another difference frequency in multiple non-radial pulsation frequency pairs.   During the outbursts in 2014/15, the light is modulated with 0.1777\,c/d, less clearly so in 2016/17. The change in mean magnitude after the outbursts is probably due to increased scattering and free-bound transitions in the subsequently dissipating ejecta.}
     \label{fig:25Ori}
\end{figure}
%MDPI: in the image, Please change hyphen (-) to minus (−) in numbers --presently not possible to do 

\subsection{$\beta$ Lyrae: A  Binary with a Hidden Component?}

A highlight binary is $\beta$\,Lyrae, which consists of a B6-8II bright giant (3\,$M_\odot$) and an invisible, more massive companion (13\,$M_\odot$) producing the primary eclipses. The bright giant loses mass to the more massive object at a rate that induces a fast period change of 19 s per year. There were no previous studies of the intrinsic variability of the $\beta$\,Lyrae system available which were credible, sufficiently continuous, and uniform, because of the day-gaps in ground-based observations, which coincided with the prevalent time-scales of the intrinsic variability in this 12.9-day orbital-period binary. 

The \br\ data extending over slightly more than 10 full orbital revolutions of the binary provided the first usable time series, reaching substantially beyond the intrinsic time scales and permitting utilisation of tools well developed for studies of variability of active galactic nuclei and quasars. Analysis of the \br\ time series shows typically three to five instability events per binary orbit, showing  a slightly stronger serial correlation than the red noise (\citet{ruc3,ruc4}). The two-parameter Damped-Random-Walk (DRW) model of the fluctuations (\citet{ke1,zu1}), characterised by the red-noise spectrum at time scales shorter than the de-correlation time scale $\tau$ and white noise at longer time scales, agrees very well with the data. 

The fluctuations are characterised by the amplitude of the stochastic signal of $1.3\,\%$, expressed relative to the maximum flux from the binary, while the de-correlation length of the random disturbances is characterised by a typical value of $\tau$ = 0.88\,days. The invisible companion is the most likely source of the instabilities. Unexpectedly, the time scale of the intrinsic variability---most likely associated with the thermal time scale of mass-transfer instabilities---appears to follow the same dependence on the mass of the accreting object as is observed for active galactic nuclei and quasi-stellar objects, which are five to nine orders of magnitude more massive than the $\beta$\,Lyrae torus-hidden component. 
%\\ {\www Text is now from Slavek}

\subsection{HD\,201433---A Rosetta-Stone SPB Star in a Multiple System}

Rotation is a still incompletely understood key process of stellar evolution \linebreak (\mbox{\citet{ae1}}). If stars locally conserved angular momentum, their cores would spin up and the surviving compact objects would spin much faster than is actually observed. This implies that present standard models are incomplete and miss essential processes and correct timescales. A first step towards solving this problem is to detect how angular momentum is distributed inside stars, as a function of various parameters including age. Prime candidates for such studies, and more easily understood than B-type stars, are subgiant and red giant stars, as they convey the rotational history of the earlier stages of evolution and pulsate with mixed p/g modes that carry information about the deep stellar interior, as is argued in \citet{ka2} and illustrated in Figure\,\ref{fig14}.

\begin{figure}[H]

     \includegraphics[width=13cm]{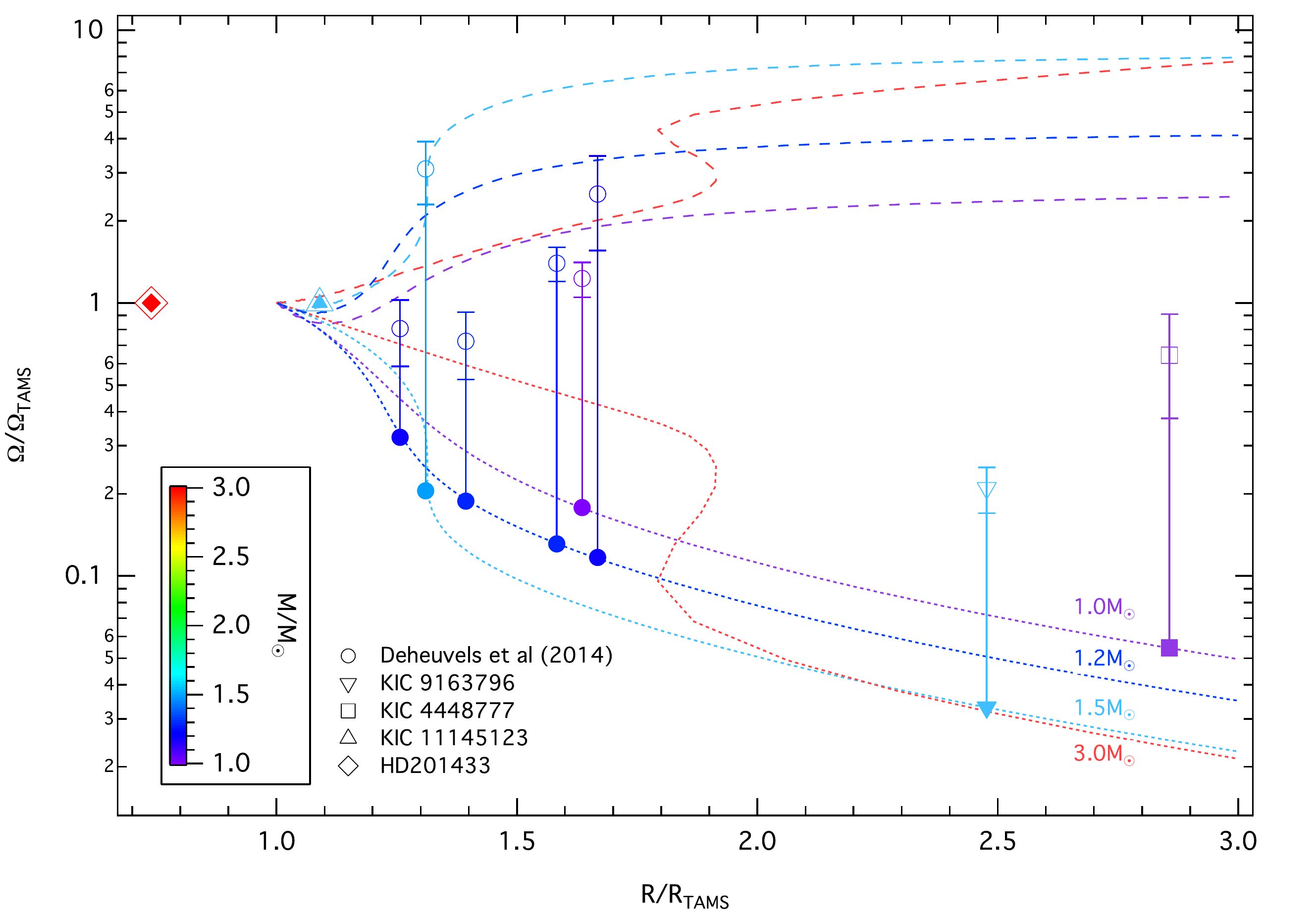}
\caption{Mean core (dashed lines) and envelope (dotted lines) rotation rate during the evolution of YREC models (from the TAMS to the RGB) with various masses (colour coded) assuming local conservation of angular momentum and rigid rotation on the main sequence. The rotation rate and stellar radius are given relative to their respective values on the TAMS. The filled symbols correspond to the relative envelope rotation rates of various stars with a given mass and radius. The core rotation rate (open symbols) is determined from this value and the observed core-to envelope rotation gradient  (Figure\,19 of \citet{ka2}).}
     \label{fig14}
\end{figure}

\textls[-15]{\bt\ observed in 2015 the SPB star HD\,201433 continuously for 156 \mbox{days  \cite{ka2}}}. The peaks in the Fourier spectrum of the \br\ observations turned out to be broader than expected, which triggered the development of a new Bayesian-based frequency determination technique with a resolution beyond the formal Rayleigh-criterion. As a proof, three rotationally split triplets are identified in the nearly half-year long \br -data, with central frequencies and splittings agreeing well with those extracted from the nearly \mbox{8 years} of \sm\ observations.

A science highlight of the HD\,201433 \br -photometry is a trend of splitting becoming more common towards longer periods, which implies a non-rigid internal rotation profile, as is elaborated in \cite{ka2}. For a detailed investigation, a dense grid of MESA \mbox{models \citep{pa1, pa2}} and their non-adiabatic pulsation modes were computed by \mbox{\citet{ka2}}. 
Using classical $\chi ^2$ techniques and other statistical methods, a representative model (3.05\,$M_{\odot}$ and 2.6\,$R_{\odot}$) was identified that reproduces best the  {observed frequencies.} 

The pulsation modes that are accessible to the seismic analysis probe the radiative envelope of HD\,201433 from the boundary of the convective core at about 0.11\,R$_{\ast}$  up to about 0.98\,R$_{\ast}$. The Bayesian analysis of various rotation profiles provides strong evidence for a slowly (292 $\pm$ 76\,d) and rigidly rotating envelope, topped by a thin and significantly more rapidly rotating surface layer, which covers about the outer 4\% of the radius (\mbox{Figure\,\ref{fig14}}).  In conclusion, \bc\ data provide strong evidence for non-rigid internal rotation in a main-sequence star, which still is rarely presented in the literature.

\subsection{The Young Star $\beta$ Pictoris and Its Exoplanetary System}   \label{bpic}

Exoplanet properties crucially depend on their host star's parameters. The $\beta$ Pic system includes a wide, dense circumstellar disk that is seen edge-on and two giant gas planets ($\beta$ Pic\,b and c) that are only grazingly eclipsing the host star.
\bc\ data have been used to search for a transiting planet.
This puts limits on the $\beta$ Pic system, as possible planets must be larger than 0.6\, (0.75, 1.0) R$_{\rm Jupiter}$ for periods of less than 5 (10, 20) days %larger than %0.75\,R$_{\rm J}$ for periods of less than 10 days and %1.0\,R$_{\rm J}$ for periods of less than 20 days 
(\citet{mol1}).

Furthermore, the predicted transit of the Hill sphere of $\beta$\,Pic\,b triggered an international observing campaign in 2017--2018 including the \bc\ nanosats. 
No dimming caused by the Hill sphere transit was observed in any of the involved photometric instruments, where the precision of the BRITE photometry would allow detection of a drop in intensity by only 0.5\% in the time of interest (\citet{ken1}). In the spectroscopic observations, some signs of the Hill sphere transit have been detected (e.g., in the Ca II H \& K lines) illustrating that the material in the planet’s Hill sphere is not sufficiently dense to dim the stellar light enough to be photometrically detected from the ground. In addition, in 1981 anomalous fluctuations of the flux coming from the $\beta$\,Pic system were originally interpreted as being caused by foreground material that transited the stellar disk. Recently, based on the observations conducted within the $\beta$\,Pic Hill sphere transit campaign, \citet{ken1} showed that this 1981 event did not originate from the transit of a circumplanetary disk.

The high-quality \bc\ photometry for $\beta$ Pictoris obtained since 2015 provided crucial constraints on the properties of the exoplanet host star itself \linebreak (\mbox{\citet{zwi5}}).~The first asteroseismic analysis using multi-color space photometry yielded a precision of 2\% in mass and radius for $\beta$ Pictoris, determined the inclination angle to be 89.1$^{\circ}$ (which agrees with the inclination angle of the disk of 88.1$^{\circ}$), and identified the \mbox{15 pulsation} frequencies as three $\ell$ = 1, six $\ell$ = 2 and six $\ell$ = 3 p-modes.

\subsection{The roAp Star $\alpha$\,Cir }   \label{s:alcir}  %%%%%%          alpha Cir

\acir \  is the brightest rapidly oscillating (roAp) star with a magnetic field. It was discovered in 1981 by \citet{dk1} and since then, many publications dealt with photometric and spectroscopic properties, including the magnetic field {(see, e.g., \citet{hol1} and \citet{we4b, we5})}.

\acir\ is a text-book illustration for an advantage of nanosatellites dedicated to photometry (e.g., \citet{we4}), as they allow one to observe stars over a long time span. Even if the accuracy of individual data points is inferior to that of larger instruments, long observations of targets result in more accurate frequency spectra. Figure\,2 of \cite{we5} presents light curves observed by five different satellites with { {apertures ranging from 3\,cm (\br -blue) to effective 10\,cm (\te) and filter bandwidths of 55\,nm and 400\,nm, centred on 425\,nm and about 800\,nm, respectively,  (see Table 1 of \cite{we5}).}} 
The larger the aperture and filter bandwidth, the more accurate the photometry, but if frequency-{\em resolution} is important, the picture changes drastically in favour of data obtained over 3 years even with a smaller aperture telescope (Figure\,\ref{fig12}). 
\begin{figure}[H]
%\vspace{-25mm}

     \includegraphics[width=12cm]{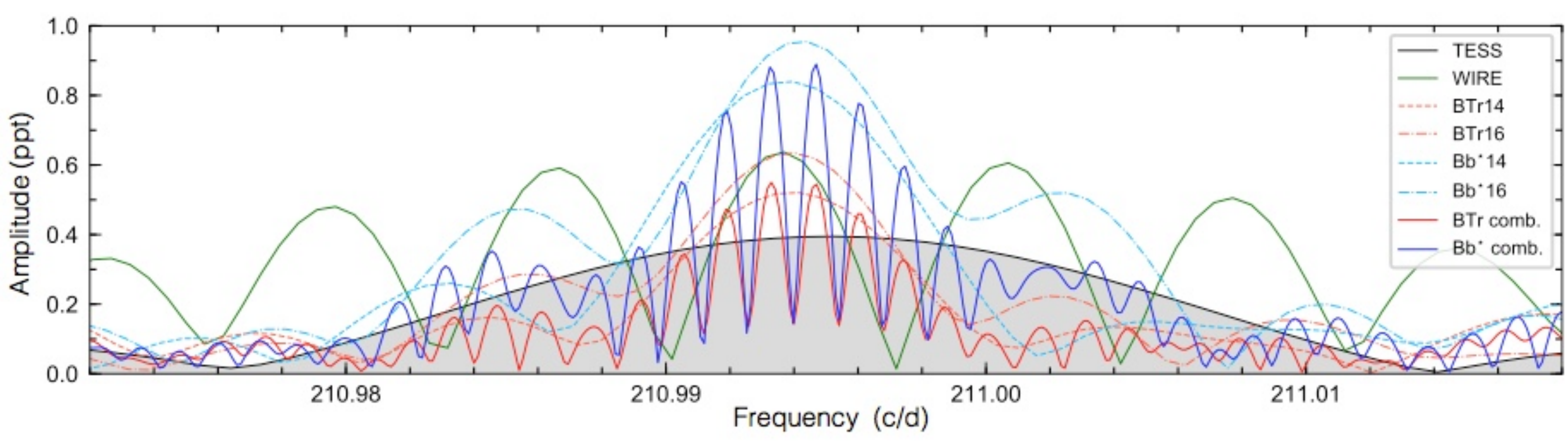}
%     \vspace{-25mm}
     \caption{Fourier amplitude spectra of the \te , \wi\ and \br\ photometry, centred on the main pulsation frequency ($f_1$) of \acir .  BTr14, BTr16, Bb$^{\star}$14, and Bb$^{\star}$16  are the red and blue BRITE data, obtained during the years 2014 and 2016, respectively. Bb$^{\star}$ represents the combined blue data obtained with \br -Austria and \br -Lem.  The  \br -blue  amplitudes are divided by two (!!) for better comparison with the other data  (adapted Figure\,6 of \citet{we5}).}
     \label{fig12}
\end{figure}

Combining the times of maximum from \br -red and \wi\ data, results in \linebreak$f_1$\,=\,210.993264(5)\,d$^{-1}$, which is, with an error in the corresponding period of 0.01\,msec, the most accurately determined dominant pulsation period of any roAp star to date. The main pulsation frequency ($f_1)$ can be identified with an $\ell=1$ mode, and two additional frequencies likely come from two consecutive radial $\ell=0$ modes \cite{we5}.

At least three surface spots can be identified for \acir  ; the \te\ data even suggest a fourth spot.  The  best-fit (minimum $\chi ^2$) set of parameters differs significantly from that inferred from the marginal distributions of the parameters, which hints at a noticeable skewness of the probability distribution of the Bayesian photometric imaging in the considered ten-dimensional configuration space. Obviously, spot latitudes are less well determined than longitudes, as expected. To our knowledge, this is the first time that Bayesian-based evidence of models differing in the number of spots has been quantitatively determined~\cite{we5}.

\subsection{$\beta$\,Cas: The First $\delta$\,Scuti Pulsator with a Dynamo Magnetic Field}

One of the cooler BRITE-Constellation targets showing pulsations and a magnetic field is the F2 type star $\beta$\,Cas, which is also one of the objects in the BRITE legacy fields (\citet{zwi6}). $\beta$\,Cas is quite an unusual star in several aspects:

\begin{itemize}
\item[(i)] It shows only two independent $\delta$\,Scuti type p-mode frequencies. As $\delta$\,Scuti stars are usually known to show up to hundreds of individual frequencies, this challenges the asteroseismic interpretation. Why only two frequencies can be detected with a total time base of over 2.5 years is still unclear. 
\item[(ii)] $\beta$\,Cas is one of the few $\delta$\,Scuti stars known to date to show a measurable magnetic field at all \cite{zwi6}. 
The three other magnetic $\delta$\,Scuti stars are HD\,188774 (\mbox{\citet{lam5}}), $\rho$\,Pup (\citet{nei6}) and HD\,41641 (\citet{tom5}).
\item[(iii)] Additionally, the magnetic field structure of $\beta$\,Cas is quite complex and almost certainly of dynamo origin. One may speculate that the presence of this dynamo field is related to the unusual lack of numerous $\delta$\,Scuti frequencies.
\end{itemize}

All this makes $\beta$\,Cas a powerful test bench for modelling of dynamo processes in thin convective envelopes of F-type stars.

\subsection{Rotation, Pulsation, Orbits and Eclipses in the Constellation of Auriga}

{ The Auriga field is an excellent example of an arrangement typically chosen for observations with \bc . One or two  key targets determine the orientation of a \br\ satellite and in the same field additional targets with a large mass-range maximise the science output.}

Rotation and pulsation periods across the Hertzsprung-Russell diagram are of top priority for understanding stellar activity as a function of time. Continuous photometry with up to three \br\  satellites was obtained for 12 targets, primarily in the Aur/Per\,I field, and subjected to a period search (\citet{kgs}). The bright active star, Capella, was found to be  constant in the red bandpass with an rms of just 1\,mmag over 176\,d, but showed a 10.1 $\pm$ 0.6\,d periodicity in the blue, which is interpreted to be the rotation period of its active and hotter secondary star (Figure\,\ref{fig16}). Its position in the Hertzsprung gap suggests ongoing changes in its internal structure. It is expected that this has a profound impact on the visible surface and can explain its fast rotation.

\begin{figure}[H]

     \includegraphics[width=8cm]{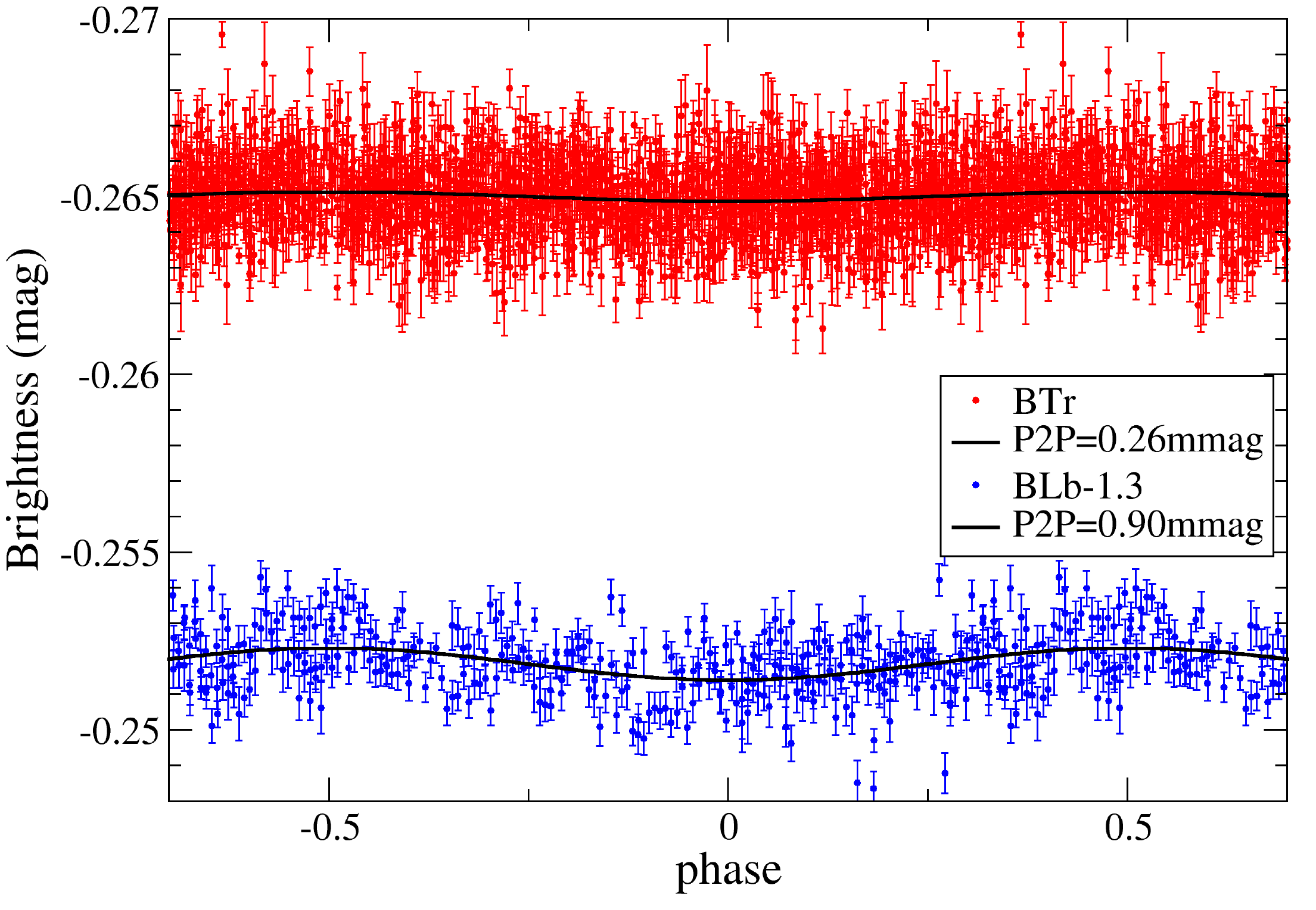}
     \caption{{Phase plots} for the red (\textbf{top}) and blue data (\textbf{bottom}) with the best-fit 10.1\,d period for Capella. The blue data are dominated by the hotter G0 component while the red data are dominated by the cooler G8 component. The rotation period of the cool component is near the orbital period of 104 days (adapted Figure\,3 of \citet{kgs}).
     }
     \label{fig16}
\end{figure}
%MDPI: in the image, Please change hyphen (-) to minus (−) in numbers --presently not possible to do 

Results for the other targets in Auriga include: 

\begin{itemize}
\item[(i)] The main pulsation period of the F0 supergiant \epsAur\ is detected by a multi-harmonic fit of the 152-day long light curve. This is noteworthy, because the RVs observed contemporaneously with the Stella spectrograph revealed a clear 68\,d period. Although the light curve showed two minima separated by 74\,d, a single period of that duration would not fit the data adequately. These RVs indicated that the (stellar) disk-integrated pulsations seem to revert when maximum or minimum light is reached, that is, the star is apparently most contracted when brightest and most expanded when faintest. 
\item[(ii)] An ingress of an eclipse of the \zAur\ binary system was covered and a precise timing for its eclipse onset derived. We obtained a possible 70 d period from the outside-eclipse light-curve fits of the proposed tidally-induced, nonradial pulsations of this ellipsoidal K4 supergiant. 
\item[(iii)] \etaAur\ was identified as an SPB star with a main period of 1.289 $\pm$ 0.001\,d. Five more periods are seen in the \br\ photometry and three of these are also seen in the RV data. The amplitude ratios as well as the phase lags between brightness and RV periods reflect those  expected from low-degree gravity modes of SPB stars. \etaAur\ is, thus, among the brightest SPB stars known. 
\item[(iv)] Rotation of the magnetic Ap star \tAur\ is easily detected by photometry and spectroscopy with a period of 3.6189 $\pm$ 0.0001\,d and 3.6177 $\pm$ 0.0006\,d, respectively. The RVs of this star show a striking non-sinusoidal shape with a large amplitude of 7\,\kms , which is likely due to the line-profile deformations from the inhomogeneous surface distribution of its chemical elements. Such a non-sinusoidal shape likely explains the small period difference and suggests that the two periods are actually \mbox{in agreement. }
\item[(v)] Photometric rotation periods are also confirmed for the magnetic Ap star \iqAur\ of 2.463\,d and for the solar-type star \kCet\ of 9.065\,d, and also for the B7 HgMn giant \bTau\ of 2.74\,d. The latter remains uncertain because it was reconstructed only with the very small amplitude of 0.54\,mmag. 
\item[(vi)] Revised orbital solutions are derived for the eclipsing SB2 binary \bAur , which replaces the initial orbit from 1948, and for the RS~CVn binary \HR\ for which a spot-corrected orbital solution was achieved. The two K giants \nAur\ and \iAur\ are found with long-term trends in both the light curve and the RVs. \nAur\ could be a long-period eccentric SB1 system with a low-mass companion for which a provisional orbital solution is predicted with a period of 20\,yr and an eccentricity of 0.7. The RV variations of the hybrid giant \iAur\ are of even lower amplitude (0.7\,\kms) but shorter period ($\approx$4\,yrs) and are more likely due to surface oscillations. Long-term brightness trends were seen for both stars and appear related with the RVs.
\end{itemize}

\subsection{Stellar Masses of Red Giants from Their Granulation Signal}     \label{redgiants}

A sample of 23 RG stars in the range 1.6 < V < 5.0 and distributed all over the sky was investigated by \citet{kal2}, and a clear granulation and/or oscillation signal was found. %(Figure\,\ref{fig17}). 
Each star was observed almost continuously by at least one of the five \br\ satellites for up to 173\,d. 

Even though plenty of information is available in the literature for these bright stars, neither surface gravity (log\,g) nor mass is sufficiently well known. Granulation and/or oscillation timescales, deduced from \bc\ observations, help to determine model-independent estimates of log\,g with two different methods \mbox{(\citet{kal3})}. Using precise radii from the literature, mostly from interferometric angular diameters and Gaia parallaxes, the mass of the stars can be estimated from log\,g, derived from \br -data, which are dominated by the granulation signal.

The stellar masses presented in Figure\,\ref{fig17} range from about 0.7 to more than $8\,M_\odot$ and have formal uncertainties of about 10\% to 20\%, which covers the observational errors as well as the known uncertainties of the used scaling relations. One might question whether simple scaling relations hold for low-mass giants with about $10\,M_\odot$ to high-mass giants with more than $200\,R_\odot$, but this is difficult to estimate due to missing independent and reliable mass estimates. Even though there might still be some unknown systematic effects in the scaling relations, they appear to be at least good enough to disentangle low-mass stars from high-mass stars. 

\begin{figure}[H]

     \includegraphics[width=13cm]{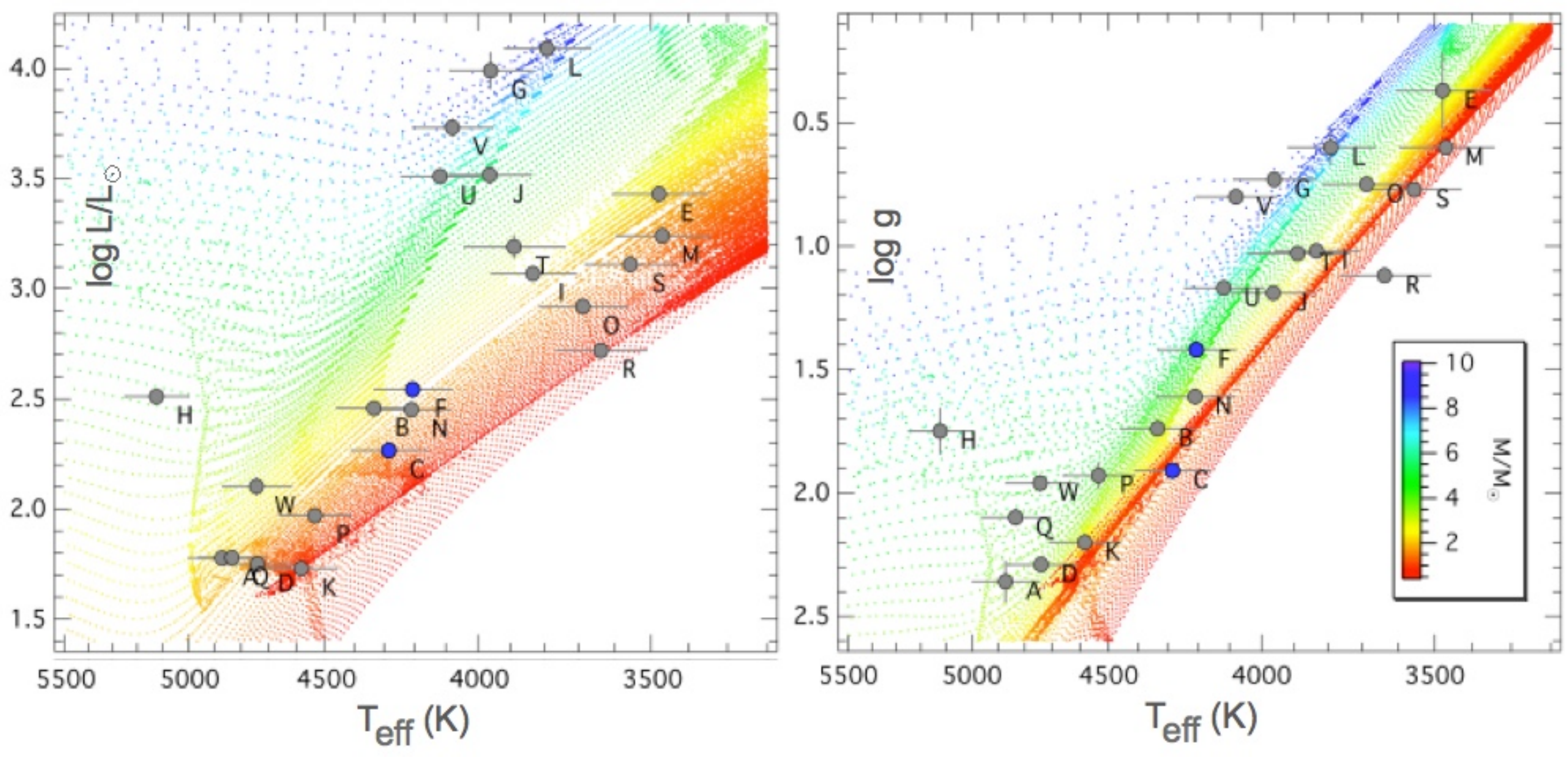}
     \caption{Hertzsprung--Russell diagram (\textbf{left}) and Kiel diagram (\textbf{right}) with red giants observed by \bc\ (grey-filled circles). The small dots show MIST stellar evolution models for solar composition with the mass colour coded. Blue-filled circles mark stars for which solar-type oscillations have been found in the \bc\ data (adapted Figure\,10 from \mbox{\citet{kal2}).}}
     \label{fig17}
\end{figure}

Comparison {{of the masses derived through the scaling relations}} with parameters from a large grid of stellar models also allows one to evaluate statistically the relative evolutionary state of the individual stars {{, that is, to distinguish low-mass red-clump stars from high-mass red giants}}. 

In recent years, the seismology of red giants has grown to become an important field in stellar astrophysics, providing the unique opportunity to probe the interior structure of evolved stars (\citet{chap1}). In general, seismic scaling relations have become indispensable for determining  mass and radius of stars with a convective envelope. 

\subsection{{{Complete Coverage of}} Nova Carinae 2018 (ASASSN-18fv)}     \label{nova} %%%%%%%        Nova

This first-time ever observation of a {\em complete} nova eruption came about by chance. The \bc\ had just monitored 18 stars continuously over several weeks in the constellation Carina, when \br -Mission-Control (MC, see Section \ref{sec32}) recognised a  sudden brightening of a field star (inserts in Figure\,\ref{fig10}). A quick search among the top sky-news announcements indicated a new star, discovered by the All-Sky Automated Survey for Supernovae (ASASSN) as ASASSN-18fv  (Figure\,\ref{fig10}). 

\begin{figure}[H]

     \includegraphics[width=12cm]{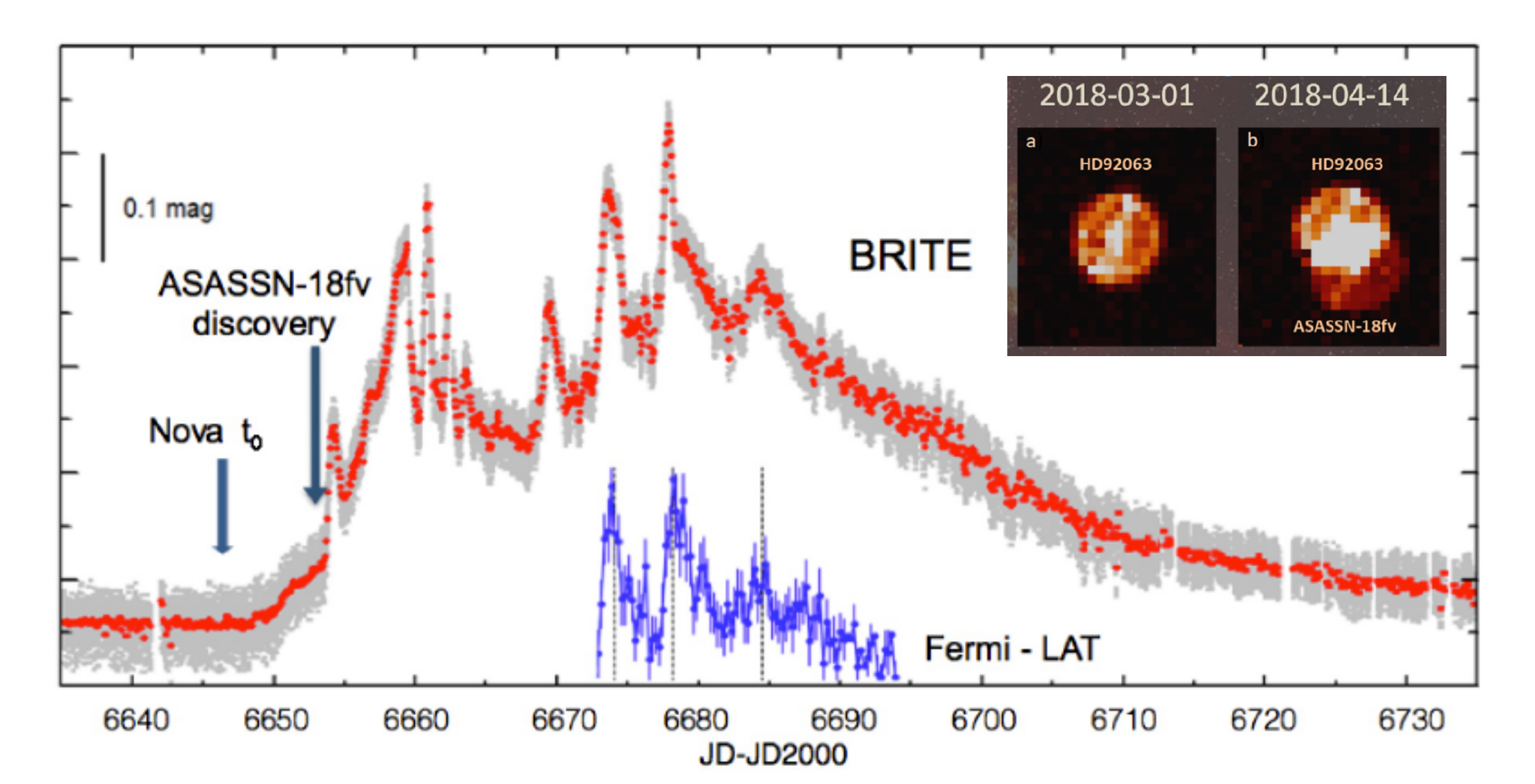}
     \caption{\bt\ photometry of Nova Carinae 2018 (grey: individual observations with red filter, red: orbit averages). Blue: Fermi-LAT (\citet{atw1}) $\gamma$-ray observations. Inserts: raster centred on HD\,92063, taken on 1 March 2018 before nova eruption (\textbf{left}), and on 14 April 2018 during nova eruption (\textbf{right}). Scale is 27$'$$'$ per pixel, and exposure times are 4\,s. }
     \label{fig10}
\end{figure}

The cooperation of \bc\ with the international community is reported, e.g., in \citet{ay1} and resulted in unprecedented simultaneous space observations in a broad wavelength range and with \br\ starting even before the \mbox{actual outburst.} 

A shock model of \citet{me1} predicts that in addition to $\gamma$-rays, the shocked gas should emit mostly in  X-rays, which will be absorbed by the dense nova ejecta ahead of the shocked gas, reprocessed to lower energies, and escape in the optical. This process indicates a source for the bolometric luminosity of the nova, in addition to the remnant nuclear burning on the white dwarf surface. Shocks occur in many transient phenomena, such as Type IIn supernovae, tidal disruption events, stellar mergers, superluminous supernovae, etc. Hence, shock interactions may contribute  substantially to the bolometric luminosities of these events, but direct observational evidence has been lacking. The \bc\ observations were unique in this context and helped to provide insights in many previously poorly observed and understood phases of novae evolution, see for example,  \citet{hou1, ay2}.

The  well-sampled \br\ light curve (Figure\,\ref{fig10}) clearly resolves  a series of distinct short-lasting flares of the order of one to two days, but which were poorly resolved from the ground. $\gamma$-rays indicate a series of flares, similar to those in the optical regime, \mbox{which suggest: }

\begin{itemize}
\item[(i)] The fact that the flares occur simultaneously in time in both \br\ bands implies that they very likely share the same origin, that is, shocks, because they power the $\gamma$-rays. Consequently, shocks are also powering some of the optical emission. 
\item[(ii)] Doubling of the luminosity of the nova during the flares, implies that the shocks power a substantial fraction of the nova luminosity. 
\item[(iii)] $\gamma$-ray and optical light curves (Figure\,\ref{fig10}) were very well sampled and indicate a time lag of approximately 5 h. This is an additional confirmation that the optical emission originates in the shocks. $\gamma$-rays escape from the shocks with little absorption, but it takes a few hours to reprocess the X-rays and to emit the energy in the optical regime, exactly as observed.
\end{itemize}
 
Fortunately, \bc\ observed this nova even before it was discovered, providing ``smoking-gun evidence'' for the shock model.

\section{Summary}
\bc\  has outlasted its minimum design-lifetime by several factors. While it is tempting to terminate the mission, it would be a real pity for humanity to do this, instead of allowing further observations to form a legacy for astronomy. The cost is truly modest compared to most other space missions, especially in relation to the valuable science that \br\ has accomplished and still can accomplish.

\bc's uniqueness lies first in the small sizes of the individual satellites that are capable of three-axis stabilization and providing a pointing stability accurate enough for astrophysical observations. Second, \bc\ is an outstanding and unique space mission because of its possibility to observe stars simultaneously in two designated pass-bands and up to 6 months contiguously. 

The big success of \bc\ is reflected---as of March 2021---in 42 peer-reviewed publications and many more conference papers that address a variety of scientific topics from the most massive stars to cool red giants and novae. Here, we have highlighted some of the key results as part of a brief overview.  
\vspace{6pt} 

\authorcontributions{All authors have contributed substantially to the work~reported. All authors have read and agreed to the published version of the manuscript.}
%For research articles with several authors, a short paragraph specifying their individual contributions must be provided. The following statements should be used ``Conceptualization, X.X. and Y.Y.; methodology, X.X.; software, X.X.; validation, X.X., Y.Y. and Z.Z.; formal analysis, X.X.; investigation, X.X.; resources, X.X.; data curation, X.X.; writing---original draft preparation, X.X.; writing---review and editing, X.X.; visualization, X.X.; supervision, X.X.; project administration, X.X.; funding acquisition, Y.Y. All authors have read and agreed to the published version of the manuscript.'', please turn to the  \href{http://img.mdpi.org/data/contributor-role-instruction.pdf}{CRediT taxonomy} for the term explanation. Authorship must be limited to those who have contributed substantially to the work~reported.

\funding{We are grateful to BEST, consisting of two voting members per satellite, supplemented by presently 15 non-voting members, who managed \bc\ successfully during nearly a decade.  
We acknowledge with thanks the variable star observations from the AAVSO International Database contributed by observers worldwide and used in this research. The authors from Poland acknowledge assistance by BRITE PMN grant 2011/01/M/ST9/05914 and GH thanks the Polish NCN for support (grant { 2015/18/A/ST9/00578}). 
AFJM and GAW acknowledge Discovery Grant support from the Natural Sciences and Engineering Research Council (NSERC) of Canada.
APo was supported by SUT grant no. 02/140/RGJ21/0012 and Statutory Activities grant no. \mbox{BK-225/RAu-11/2021.}}

%%Open Access Funding by the University of Vienna
%Please add: ``This research received no external funding'' or ``This research was funded by NAME OF FUNDER grant number XXX.'' and  and ``The APC was funded by XXX''. Check carefully that the details given are accurate and use the standard spelling of funding agency names at \url{https://search.crossref.org/funding}, any errors may affect your future funding.

\institutionalreview{Not applicable.}
%In this section, please add the Institutional Review Board Statement and approval number for studies involving humans or animals. Please note that the Editorial Office might ask you for further information. Please add “The study was conducted according to the guidelines of the Declaration of Helsinki, and approved by the Institutional Review Board (or Ethics Committee) of NAME OF INSTITUTE (protocol code XXX and date of approval).” OR “Ethical review and approval were waived for this study, due to REASON (please provide a detailed justification).” OR “Not applicable” for studies not involving humans or animals. You might also choose to ex-clude this statement if the study did not involve humans or animals.

\informedconsent{Not applicable.}
%Any research article describing a study involving humans should contain this statement. Please add “Informed consent was obtained from all subjects involved in the study.” OR “Patient con-sent was waived due to REASON (please provide a detailed justification).” OR “Not applicable” for studies not involving humans. You might also choose to exclude this statement if the study did not involve humans.

%Written informed consent for publication must be obtained from participating patients who can be identified (including by the patients themselves). Please state “Written informed consent has been obtained from the patient(s) to publish this paper” if applicable.

\dataavailability{Data access is provided by  \url{https://brite.camk.edu.pl/pub/index.html}.}
%In this section, please provide details regarding where data supporting reported results can be found, including links to publicly archived datasets analyzed or generated during the study. Please refer to suggested Data Availability Statements in section “MDPI Research Data Policies” at \href{https://www.mdpi.com/ethics}{https://www.mdpi.com/ethics}. You might choose to exclude this statement if the study did not report any data.

%\acknowledgments{ 
%}

\conflictsofinterest{The authors declare no conflict of interest.}
%Declare conflicts of interest or state ``The authors declare no conflict of interest.'' Authors must identify and declare any personal circumstances or interest that may be perceived as inappropriately influencing the representation or interpretation of reported research results. Any role of the funders in the design of the study; in the collection, analyses or interpretation of data; in the writing of the manuscript, or in the decision to publish the results must be declared in this section. If there is no role, please state ``The funders had no role in the design of the study; in the collection, analyses, or interpretation of data; in the writing of the manuscript, or in the decision to publish the~results''.

\end{paracol}

\reftitle{References}

\end{document}